\documentclass{article}

 \usepackage[main, final]{neurips_2025}
 
\usepackage[utf8]{inputenc} 
\usepackage[T1]{fontenc}    
\usepackage{hyperref}       

\usepackage{url}            
\usepackage{booktabs}       
\usepackage{amsfonts}       
\usepackage{nicefrac}       
\usepackage{microtype}      
\usepackage{xcolor}         
\usepackage{wrapfig}        
\usepackage{graphicx}       
\usepackage{amsmath}
\usepackage{multirow}
\usepackage[table]{xcolor}
\usepackage{stfloats}
\usepackage{float}
\usepackage{enumitem}
\usepackage{makecell}

\usepackage{caption}
\usepackage{subcaption}

\definecolor{best}{rgb}{0.8, 0.9, 1.0}
\definecolor{2ndbest}{rgb}{0.9, 1.0, 0.8}

\newcommand{\best}[1]{\colorbox{best}{\textbf{#1}}}
\newcommand{\secondbest}[1]{\colorbox{2ndbest}{#1}}

\usepackage{algorithm}
\usepackage{algcompatible}
\algnewcommand\algorithmicreturn{\textbf{return}}
\algnewcommand\RETURN{\State \algorithmicreturn}%

\usepackage{lipsum}

\makeatletter
\usepackage{xspace}
\def\@onedot{\ifx\@let@token.\else.\null\fi\xspace}
\DeclareRobustCommand\onedot{\futurelet\@let@token\@onedot}

\newcommand{\eqnref}[1]{Eq\onedot~\eqref{#1}}
\newcommand{\figref}[1]{Fig\onedot~\ref{#1}}
\newcommand{\algoref}[1]{Alg\onedot~\ref{#1}}
\newcommand{\secref}[1]{Section~\ref{#1}}
\newcommand{\tabref}[1]{Tab\onedot~\ref{#1}}
\newcommand{\appref}[1]{Appendix~\ref{#1}}
\newcommand{\triangleqdef}{\mathrel{\stackrel{\triangle}{=}}}

\def\eg{\emph{e.g}\onedot}
\def\ie{\emph{i.e}\onedot}
\def\iid{i.i.d\onedot}
\def\x{{\mathbf x}}
\def\y{{\mathbf y}}

\title{Time-Embedded Algorithm Unrolling for Computational MRI}

\author{
  Junno Yun \\
  University of Minnesota\\
  \texttt{yun00049@umn.edu} \\
  \And
  Ya\c{s}ar Utku Al\c{c}alar \\
  University of Minnesota\\
  \texttt{alcal029@umn.edu} \\
  \And
  Mehmet Ak\c{c}akaya\thanks{Corresponding Author} \\
  University of Minnesota\\
  \texttt{akcakaya@umn.edu}
}

\begin{document}

\maketitle
\begin{abstract}
Algorithm unrolling methods have proven powerful for solving the regularized least squares problem in computational magnetic resonance imaging (MRI). These approaches unfold an iterative algorithm with a fixed number of iterations, typically alternating between a neural network-based proximal operator for regularization, a data fidelity operation and auxiliary updates with learnable parameters. While the connection to optimization methods dictate that the proximal operator network should be shared across unrolls, this can introduce artifacts or blurring. Heuristically, practitioners have shown that using distinct networks may be beneficial, but this significantly increases the number of learnable parameters, making it challenging to prevent overfitting. To address these shortcomings, by taking inspirations from proximal operators with varying thresholds in approximate message passing (AMP) and the success of time-embedding in diffusion models, we propose a time-embedded algorithm unrolling scheme for inverse problems. Specifically, we introduce a novel perspective on the iteration-dependent proximal operation in vector AMP (VAMP) and the subsequent Onsager correction in the context of algorithm unrolling, framing them as a time-embedded neural network. Similarly, the scalar weights in the data fidelity operation and its associated Onsager correction are cast as time-dependent learnable parameters. Our extensive experiments on the fastMRI dataset, spanning various acceleration rates and datasets, demonstrate that our method effectively reduces aliasing artifacts and mitigates noise amplification, achieving state-of-the-art performance. Furthermore, we show that our time-embedding strategy extends to existing algorithm unrolling approaches, enhancing reconstruction quality without increasing the computational complexity significantly. Code available at \url{https://github.com/JN-Yun/TE-Unrolling-MRI}.

\end{abstract}

\section{Introduction} \label{sec:intro}

Algorithm unrolling/unfolding has emerged as an effective method for addressing inverse problems in computational MRI~\cite{liang2020deep, hammernik2018VarNet, aggarwal2019MoDL, hammernik2023SPM, knoll2020deep-survey, ramzi2022nc-pdnet, mardani2018neural, schlemper2018deep, yaman2020SSDU}. In this framework, traditional iterative optimization problems are unrolled for a fixed number of steps, with the network alternating between enforcing data fidelity based on the known physics-based forward operator and applying implicit regularization via a neural network based proximal operator. This unrolled network is trained end-to-end to jointly optimize the weight(s) for data fidelity and the neural network parameters for the proximal operator. Several different optimization methods have been explored for algorithm unrolling in MRI~\cite{liang2020deep, knoll2020deep-survey, fessler2020SPM, hyun2018deep, yaman2022zeroshot, hosseini2020dense}, including gradient descent (GD)~\cite{hammernik2018VarNet}, proximal gradient descent (PGD)~\cite{schlemper2018deep, zhang2018ista-net, hosseini2020dense, mardani2018neural}, variable splitting with quadratic penalty (VSQP)~\cite{aggarwal2019MoDL, duan2019vs-net, yaman2020SSDU} and alternating direction method of multipliers (ADMM)~\cite{sun2016deep, yang2020admm_csnet}, among others~\cite{ramzi2022nc-pdnet, adler2018learned_primal-dual}.

\begin{figure}[t]
    \centering
    \includegraphics[width=\linewidth]{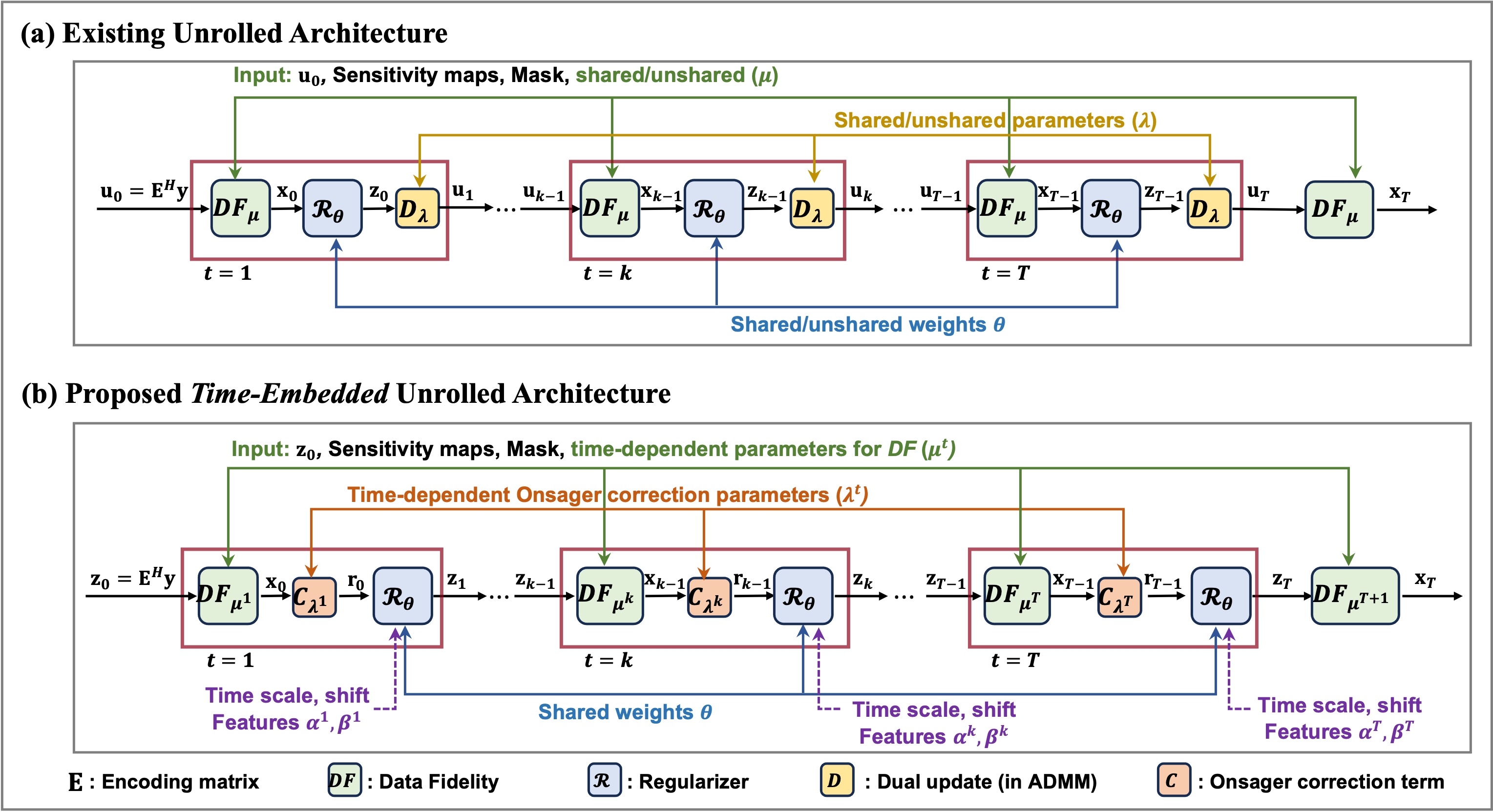} 
    \vspace{-0.3cm}
    \caption{Descriptions of (a) the existing unrolled architecture and (b) the proposed time-embedded unrolled architecture.}
    \label{fig:methods}
    \vspace{-0.4cm}
\end{figure}

Beyond the choice of optimization framework, another critical design decision in algorithm unrolling is whether to share the proximal operator for the learned regularizer across unrolls. While theoretical connections to optimization theory suggest that the proximal operator should remain fixed to maintain consistency with traditional methods~\cite{aggarwal2019MoDL,monga2021algorithm}, this may lead to unwanted artifacts. To address this, many practitioners instead allow the proximal operator to vary across iterations, effectively using distinct networks to learn iteration-specific regularization~\cite{hammernik2018VarNet,knoll2020fastMRIchallenge-1,muckley2021fastMRIchallenge-2}. This empirical strategy often enhances reconstruction quality but comes with practical trade-offs, such as larger number of trainable parameters which can heighten the risk of over-fitting, especially for applications with \emph{limited training data}~\cite{akcakaya2019RAKI,yaman2022zeroshot,darestani2021accelerated}. Such limited data settings, which are the focus of this study, are especially important in many translational applications, where new sequences are being implemented or higher resolutions are being pursued, as it is often not feasible to curate databases with thousands of slices.

A related perspective on iterative reconstruction emerges from approximate message passing (AMP) methods, which have been developed as an iterative Bayesian estimator for recovering a sparse signal for certain classes of measurement matrices in the context of compressed sensing~\cite{donoho2010message, donoho2009message}. AMP adapts the proximal operator at each iteration based on the prior distribution and includes an \textit{Onsager correction} term to stabilize the process and accelerate signal recovery. A notable extension, vector AMP (VAMP)~\cite{rangan2019vector}, improves this approach by introducing vector-valued variable nodes, and estimating each node using Minimum Mean Square Error (MMSE) and Linear MMSE (LMMSE) estimators while preserving the properties of AMP. This enables VAMP to remain effective for a broader class of measurement matrices, extending the applicability of AMP to a broader class of measurement matrices. 

Similar to the iteration-dependent proximal operator in AMP methods, a time-dependent denoiser has shown to be highly effective in the context of diffusion models~\cite{ho2020ddpm,song2021scoreSDE,dhariwal2021beatGANs,ho2021classifierfree,jalal2021robust}. In diffusion-based approaches, the denoiser adapts dynamically at each time step to better preserve structure and enhance signal recovery~\cite{song2019generative,ho2020ddpm}. This time-dependent adjustment has been shown to outperform static denoisers, particularly in tasks requiring high fidelity and sharpness~\cite{ho2020ddpm,dhariwal2021beatGANs}, as it allows the model to better handle varying noise levels throughout the diffusion process.

Building on these principles from AMP methods and diffusion models, we propose a novel time-embedded unrolling of optimization algorithms, theoretically motivated from VAMP. Our main contributions are: 
\begin{itemize}[leftmargin=*, itemsep=0em, topsep=0pt]
\item We introduce time (or iteration)-dependent unrolling of optimization algorithms by incorporating time-information into the proximal operator, theoretically motivated by VAMP formalism. To the best of our knowledge, our approach is the first attempt to bring in the time information into the algorithm unrolling to further improve performance with minimal increase in computational complexity. 
\item Our method also learns the guidance scale (\ie, the data fidelity weight in our case) in a time-dependent manner during training, which is a major deviation from commonly used guidance methods in diffusion models~\cite{dhariwal2021beatGANs,ho2021classifierfree}.
\item We demonstrate that our time-embedding strategy can be extended to various optimization algorithms, such as VSQP and ADMM, and applied to different neural network architectures for the proximal operator.
\item  We showcase the efficacy of incorporating the time information to the unrolling process through both quantitative and qualitative assessments on fastMRI dataset~\cite{knoll2020fastmri_dataset-journal,zbontar2019fastmri_dataset-arXiv}. Our approach performs on par with methods that use distinct proximal operator weights, which has substantially more learnable parameters and may face performance decrease in small training database such as ours. Furthermore, our method consistently outperforms the baseline shared-regularizer approach across various acceleration rates, producing artifact-free reconstructions with minimal processing overhead.
\end{itemize}

\section{Background and Related Work} \label{sec:background}

\subsection{Inverse Problems in Computational MRI}
A canonical problem in computational MRI is to recover an image $\x \in \mathbb{C}^N$ from noisy sub-sampled measurements $\y_{\Omega} \in \mathbb{C}^M$. The forward model in this case is given as 
\begin{equation}
    \y_{\Omega} = \mathbf{E}_{\Omega}\x + \mathbf{n}, \label{eq:problem}
\end{equation}
where $\mathbf{E}_{\Omega} \in \mathbb{C}^{M \times N}$ is a known encoding matrix that samples the Fourier domain (\ie k-space) locations specified by $\Omega$, and includes coil sensitivities, and $\mathbf{n} \in \mathbb{C}^M$ is \iid Gaussian measurement noise. The inverse problem corresponding to \eqnref{eq:problem} is typically ill-conditioned~\cite{akcakaya2022reconbook,lustig2007sparse, Akcakaya2011pvmra, akccakaya2011low, akccakaya2010compressed}, necessitating additional regularization to be incorporated into the objective function~\cite{hammernik2023SPM}: 
\begin{equation}
    \arg\min_\mathbf{x} \left\| \mathbf{y}_{\Omega} - \mathbf{E}_{\Omega} \mathbf{x} \right\|_2^2 + \mathcal{R}(\mathbf{x}), \label{eq:problem2}
\end{equation}
where the first term ensures data fidelity with the acquired measurements and $\mathcal{R}(\cdot)$ is a regularizer. 

\subsection{Algorithm Unrolling} \label{sec:2_2}
The optimization problem in \eqnref{eq:problem2} can be solved using various methods~\cite{fessler2020SPM}, including VSQP~\cite{afonso2010fast} and ADMM~\cite{boyd2011admm}, all of which have been explored in algorithm unrolling. The unrolled network iterates between data fidelity and regularization, with the latter implicitly enforced via a neural network, as illustrated in \figref{fig:methods}(a). VSQP unrolling~\cite{aggarwal2019MoDL, demirel2021EMBC_20fold_7TfMRI, duan2019vs-net, yaman2021_3D-LGE, yaman2020SSDU,yaman2022mmssdu} solves \eqnref{eq:problem2} via:
\begin{align}
    \mathbf{x}^{t} &= \left( \mathbf{E}^H_{\Omega} \mathbf{E}_{\Omega} + \mu \mathbf{I} \right)^{-1} \left( \mathbf{E}^H_{\Omega} \mathbf{y}_{\Omega} + \mu \mathbf{z}^{t}  \right), \label{eq:vs_x_sol} \\
   \mathbf{z}^{t+1} &= \arg\min_\mathbf{\mathbf{z}} \frac{1}{2} \left\| \mathbf{x}^t - \mathbf{z} \right\|_2^2 + \mathcal{R}(\mathbf{z}) \triangleqdef \text{Prox}_{\mathcal{R}}(\mathbf{x}^{t}), 
   \label{eq:vs_z_sol}
\end{align}
where the data fidelity parameter $\mu$ is learnable, and $\text{Prox}_{\mathcal{R}}(\cdot)$ is learned implicitly via a neural network. While \eqnref{eq:vs_x_sol} has a closed-form solution, it is numerically solved using the CG method~\cite{aggarwal2019MoDL}. In contrast, ADMM is a commonly used optimization approach with better convergence than VSQP~\cite{boyd2011admm}, owing to an additional Lagrangian update, and has been popular in algorithm unrolling~\cite{yang2020admm_csnet, sun2016deep, gu2022revisiting, demirel2023SIIM}:
\begin{align}
    \mathbf{x}^{t+1} &= \left( \mathbf{E}^H_{\Omega} \mathbf{E}_{\Omega} + \mu \mathbf{I} \right)^{-1} \left( \mathbf{E}^H_{\Omega} \mathbf{y}_{\Omega} + \mu \left( \mathbf{z}^{t} - \mathbf{u}^t \right) \right), \label{eq:admm_x_sol} \\
    \mathbf{z}^{t+1} &= \text{Prox}_{\mathcal{R}}(\mathbf{x}^{t+1} + \mathbf{u}^t), \label{eq:admm_z_sol} \\
    \mathbf{u}^{t+1} &= \mathbf{u}^{t} + \lambda(\mathbf{x}^{t+1} - \mathbf{z}^{t+1}) ,\label{eq:admm_u_update}
\end{align}
where $\mathcal{R}(\cdot)$, $\mu$ and $\lambda$ are learnable~\cite{yang2020admm_csnet, sun2016deep}.

Although in all of the cases, optimization theory~\cite{fessler2020SPM} dictates that $\mathcal{R}(\cdot)$ in \eqnref{eq:problem2} should be fixed across unrolls, researchers heuristically realized enabling $\mathcal{R}(\cdot)$ to change across iterations yields better reconstructions~\cite{knoll2020fastMRIchallenge-1,muckley2021fastMRIchallenge-2}. However, this increases the number of trainable parameters, and the risk of overfitting, particularly in data-limited settings.

\subsection{Approximate Message Passing} \label{sec:2_3}
AMP~\cite{donoho2010message, donoho2009message} provides an alternative approach to solving \eqnref{eq:problem} when $\mathbf{E}$ is a large i.i.d\onedot (sub-Gaussian) matrix. The AMP algorithm uses an \textit{iteration-dependent} proximal operator and an \textit{Onsager correction} term, which together enable faster convergence compared to PGD~\cite{daubechies2004iterative}. At iteration $t$, AMP applies the proximal operator with threshold proportional to \( \sigma^t \) that represents an estimate of the mean squared error of the current estimate. However, when the measurement matrix deviates from the i.i.d. sub-Gaussian regime, AMP methods often fail to converge~\cite{rangan2019vector}.

Vector AMP (VAMP) algorithm~\cite{rangan2019vector} is an alternative, offering convergence in the large \( N \) limit for a broader class of matrices \( \mathbf{E} \). It extends the AMP framework to vector-valued nodes~\cite{rangan2011generalized, opper2005expectation, seeger2005expectation}, and has connections to the ADMM algorithm~\cite{manoel2018approximate, opper2005expectation}, while preserving the desirable properties of AMP. These vector-valued operations lead to a data fidelity operation based on linear MMSE estimation, and its associated Onsager correction as:
\begin{align}
    {\bf x}^t &=  (\mathbf{E}^H \mathbf{E} + \mu_x^t \mathbf{I})^{-1} (\mathbf{E}^H \mathbf{y} + \mu_x^t \mathbf{r}^t), \label{eq:lmmse_estimator1} \\
    \upsilon_x^t &= \frac{1}{N} \mathrm{Tr} \left[(\mathbf{E}^H \mathbf{E} + \mu_x^t \mathbf{I})^{-1} \right]; \: \mu_z^t= \frac{1}{\upsilon_x^{t}} - \mu_x^{t}; \:  
   {\bf u}^t =  \left (\frac{\mathbf{x}^{t}}{\upsilon_x^{t}} - \mu_x^{t} \mathbf{r^t} \right) / \mu_z^t  \label{eq:lmsse_Onsager}
\end{align}
followed by the proximal operator/denoising step with its Onsager correction:
\begin{align}
    \mathbf{z}^{t} &= \text{Prox}_{\mathcal{R}_{\mu_z^{t}}}(\mathbf{u}^t), \label{eq:vamp1} \\
     \upsilon_z^{t} &= \frac{1}{\mu_z^{t}}\left\langle \nabla \text{Prox}_{\mathcal{R}_{\mu_z^{t}}}(\mathbf{u}^t) \right\rangle; \: \mu_x^{t+1} = \frac{1}{\upsilon_z^{t}} - \mu_z^{t}; 
    \: \mathbf{r}^{t+1} = \left (\frac{\mathbf{z}^{t}}{\upsilon_z^{t}} - \mu_z^{t} \mathbf{u}^t \right) / \mu_x^{t+1}. \label{eq:vamp2}
\end{align}
Notably, both data fidelity and denoising steps have parameters, $\mu_x^t, \mu_z^t,$ which are functions of the iteration number.

We note that AMP and its variants have also been explored in the context of algorithm unrolling. In~\cite{borgerding2016onsager} and~\cite{borgerding2017amp}, $\mathbf{E}$ and $\mathbf{E}^H$ are reparameterized with tunable parameters as $\beta^t\mathbf{E}$ and $\mathbf{E}^H \mathbf{C}^t$, where $\beta^t$ and $\mathbf{C}^t$ are trainable across unrollings via neural networks. This reparameterization influences the Onsager correction term and the denoising threshold, improving the robustness of $\mathbf{E}$. Subsequent studies have explored training distinct, \ie unshared in our previous terminology, proximal operators~\cite{metzler2017learned, karan2024unrolled} over iterations using neural networks, rather than reparameterizing the matrix $\mathbf{E}$ and $\mathbf{E}^H$. Other studies have also explored  training both the system matrix and proximal operators using neural networks~\cite{zhang2020amp, karan2024unrolled} across iterations.  
\subsection{Time-Embedding in Neural Networks} \label{sec:2_4}
Time-dependent processing plays a crucial role in diffusion models as well~\cite{ho2020ddpm,song2021scoreSDE,dhariwal2021beatGANs,ho2021classifierfree}. In this context, information about the current diffusion step is encoded to guide the CNN model to capture sequential relationships effectively and to reverse the noise process efficiently.

\begin{figure}[t]
  \centering
   \includegraphics[width=0.99\linewidth]{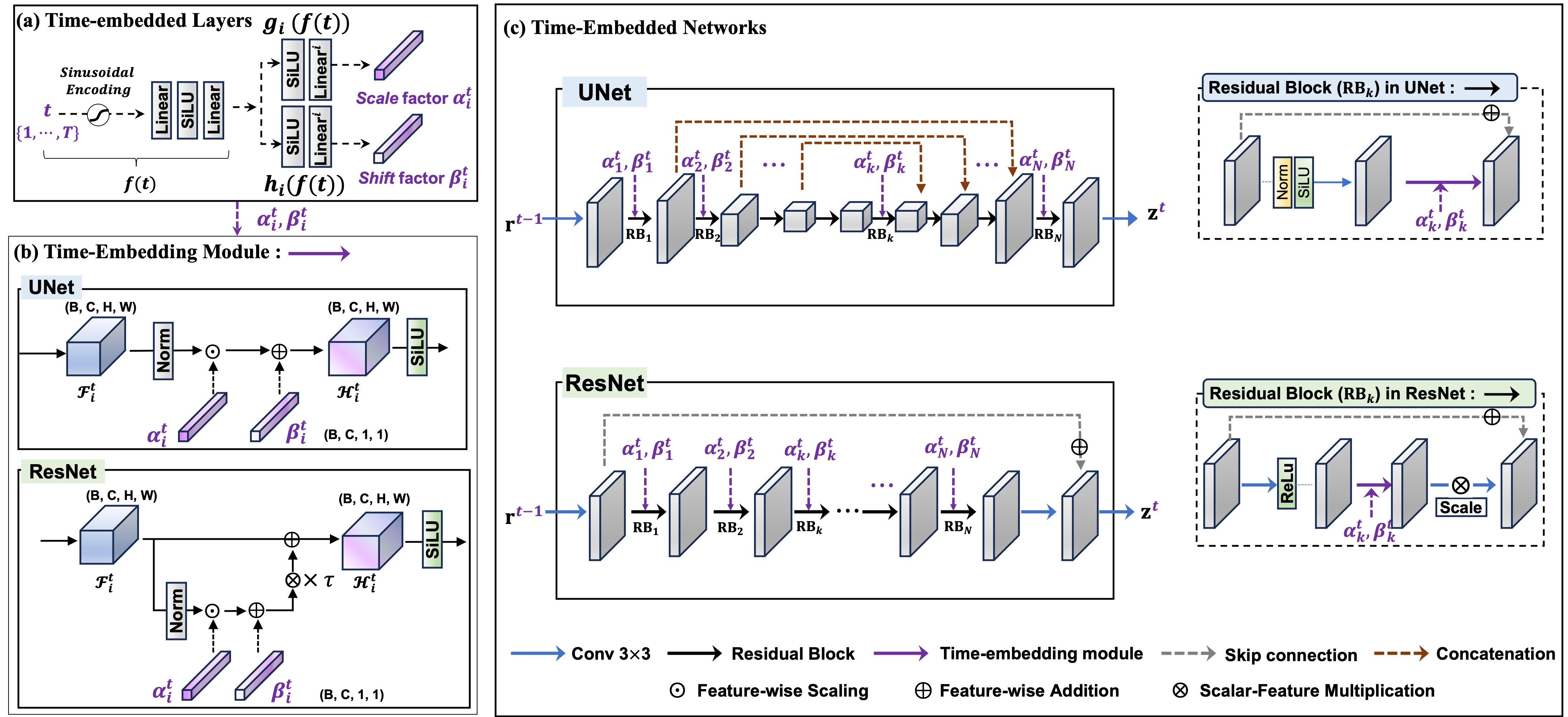}
   \caption{Illustrations of (a) Positional Encoder generating time-dependent scaling ($\alpha^t_i$) and shift ($\beta^t_i$) features, (b) Time-embedding module for ResNet and U-Net, and (c) Architectures of ResNet and U-Net showing how time-embedded features are applied.}
   \vspace{-0.2cm}
   \label{fig:time_emb_networks}
    \vspace{-0.1cm}
\end{figure}

Feature-wise linear modulation (FiLM)~\cite{perez2018film} is widely utilized for transforming inputs with time-embedded features, as illustrated in \figref{fig:time_emb_networks} (a) and (b). The time information features are obtained through a sinusoidal encoder~\cite{vaswani2017attention}, followed by a learned function \( f(\cdot) \). Subsequently, the functions \( g_i \) and \( h_i \) are adaptively learned to generate \( \alpha_i^t \) and \( \beta_i^t \), respectively as:
\begin{equation}
    \alpha_i^t = g_i(f(t)); \qquad \beta_i^t = h_i(f(t)), \label{eq:FiLM-1}
\end{equation}
where \( \alpha_i^t \) and \( \beta_i^t \) modulate the \( i^{th} \) features $ \mathcal{F}_i^t $ of CNNs at the \( t^{th} \) iteration using FiLM, which applies scaling and shifting transformations using \( \alpha_i^t \) and \( \beta_i^t \) respectively. Moreover,~\cite{dhariwal2021beatGANs} demonstrates that combining group normalization~\cite{wu2018group_norm} with the FiLM approach enhances the efficacy of time-embedded features, leading to improved model performance in diffusion model as follows:
\begin{equation}
    \mathcal{H}_i^t = \alpha_i^t \odot \text{GroupNorm}(\mathcal{F}_i^t) \oplus \beta_i^t, \label{eq:FiLM-2}
\end{equation}
where \( \mathcal{H}_i^t \) are features conditioned by time-embedded layers, \( \odot \) is feature-wise multiplication, \( \oplus \) is feature-wise addition. Each feature map in the network is modulated independently by \( \alpha_i^t \) and \( \beta_i^t \). For example, \( \mathcal{H}_i^t \) is passed onto the next block as input \( \mathcal{F}_{i+1}^t \) and modulated by the next time-embedded scaling and shifting factors \( \alpha_{i+1}^t \) and \( \beta_{i+1}^t \).

\section{Methodology}
\label{sec:Methodology}

\subsection{Proposed Time-Embedding Strategies for Algorithm Unrolling} \label{sec:3_1}
Building on \secref{sec:2_3} and \secref{sec:2_4}, we propose a time-embedding framework for algorithm unrolling. Our time-dependent proximal operator is inspired by VAMP, and is implemented as a CNN with the time-embedding techniques from \secref{sec:2_4}. This enables the proximal operator to exploit temporal dependencies across iterations, adapting its behavior dynamically, similar to the denoising in diffusion models. 
We note that, while the time step $t$ explicitly models the noise level in the given stage in diffusion models, in our case with algorithm unrolling, it implicitly modulates the proximal operator’s behavior across \emph{iterations} by capturing the evolving distribution of intermediate features, similar to the effect of the Onsager correction in VAMP.

\paragraph{Time-embedding in proximal operators} We first consider the proximal operator step of VAMP in \eqnref{eq:vamp1} and its corresponding Onsager corrections in \eqnref{eq:vamp2}. In the context of algorithm unrolling, these suggest that the learned proximal operator should be time-dependent, but with a well-defined time schedule. Furthermore, the Onsager corrections in \eqnref{eq:vamp2} are also functions of time in the original VAMP setting, and ${\bf r}^{t+1}$ in \eqnref{eq:vamp2} is effectively a function of ${\bf u}^t$ in \eqnref{eq:vamp1} and $t$. To this end, in an unrolled network setting, where the intermediate parameters can also be learned, we propose to model this relationship directly with a time-embedded neural network. In other words, a time-embedded neural network is used to model all steps in \eqref{eq:vamp1}-\eqref{eq:vamp2}, effectively capturing both the time-dependent denoising and the associated Onsager corrections implicitly to map ${\bf r}^{t+1}$ from ${\bf u}^t$. We represent this relationship as:
\begin{equation}
    \mathbf{r}^{t+1} = \text{prox}_\mathcal{R}(\mathbf{u^{t+1}}, \alpha^t, \beta^t, t),
\end{equation}
where $\alpha^t$ and $\beta^t$ capture the time-embedding information as described in \secref{sec:2_4}.

\begin{wrapfigure}[12]{r}{0.55\textwidth}
\vspace{-2.8em}
\begin{minipage}{0.55\textwidth}
    \input{algs_and_tables/alg_proposed}
\end{minipage}
\end{wrapfigure}

\paragraph{Time-embedding for data fidelity} The data fidelity term in \eqnref{eq:lmmse_estimator1} is of the same form as the data fidelity term in \eqnref{eq:vs_x_sol} in VSQP, with the notable distinction that the quadratic penalty \( \mu^t \) evolves in a time-dependent manner in the former. Thus, we implement \( \mu^t \) as a time-dependent learnable parameter. Furthermore, the Onsager correction in \eqref{eq:lmsse_Onsager} can be written as:
\begin{equation}
    {\bf u}^t = {\bf x}^t + \rho^t ({\bf x}^t - {\bf r}^t),
\end{equation}
where $\rho^t = \frac{\mu_x^t}{1/\upsilon_x^t - \mu_x^t}$. Thus, in the algorithm unrolling framework, the scalars in \eqnref{eq:lmsse_Onsager} can be replaced with a time-dependent learnable parameter $\rho^t$ for a learned Onsager correction term. Thus, the full time-embedded unrolled network is summarized in \algoref{algo:alg_proposed}.

\begin{table}[t]
    \caption{$\spadesuit$: Shared $\mathcal{R}(\cdot)$ weights, $\clubsuit$: Unshared $\mathcal{R}(\cdot)$ weights. Quantitative results are reported using \emph{limited data} on the coronal PD, coronal PD-FS, and axial T2 datasets, with equispaced undersampling patterns at acceleration rates $R=4$, $6$, and $8$. The \colorbox{best}{{\bf best}} and \colorbox{2ndbest}{second-best} results for each architecture are highlighted.}
    \label{tab:table_quant_main}
    \centering
    \small
    \setlength{\tabcolsep}{0.4pt}
    \renewcommand{\arraystretch}{0.4}
    \begin{tabular}{@{}p{0.4cm}rccccc|cc|cccc|cc@{}}
        \toprule
         &  &  & \multicolumn{6}{c}{{\bf U-Net}} & \multicolumn{6}{c}{{\bf ResNet}}\\
         \cmidrule(lr){4-9} \cmidrule(lr){10-15}
        \textbf{ } & \textbf{R} & & {\tiny \textbf{VSQP ($\spadesuit$)}} & {\tiny \textbf{VSQP ($\clubsuit$)}} & {\tiny \textbf{ADMM ($\spadesuit$)}} & {\tiny \textbf{ADMM ($\clubsuit$)}} & {\tiny \textbf{\shortstack{Ours \\ (5 unrolls)}}} & {\tiny \textbf{\shortstack{Ours \\ (10 unrolls)}}} & {\tiny \textbf{VSQP ($\spadesuit$)}} & {\tiny \textbf{VSQP ($\clubsuit$)}} & {\tiny \textbf{ADMM ($\spadesuit$)}} & {\tiny \textbf{ADMM ($\clubsuit$)}} & {\tiny \textbf{\shortstack{Ours \\ (5 unrolls)}}} & {\tiny \textbf{\shortstack{Ours \\ (10 unrolls)}}} \\ \midrule
        \multirow{6}{*}{\rotatebox{90}{\parbox{1.8cm}{\centering \textbf{Coronal PD}}}}
        & \multirow{2}{*}{\tiny $\times 4$} & {\tiny PSNR$\uparrow$}
        & 40.50 & 40.31 & 40.76 & 40.51 & \secondbest{40.94} & \best{40.99} 
        & 41.11 & 40.99 & 41.27 & 41.11 & \secondbest{41.41} & \best{41.43}  \\

        &  & {\tiny SSIM$\uparrow$} 
        & 0.962 & 0.960 & 0.964 & 0.963 & \secondbest{0.964} & \best{0.964}
        & 0.965 & 0.963 & 0.964 & 0.964& \best{0.966} & \secondbest{0.965} \\
        
        \arrayrulecolor{gray} \cmidrule(lr){2-15}
        
        & \multirow{2}{*}{\tiny $\times 6$} & {\tiny PSNR$\uparrow$} 
        & 38.12 & 38.02 & 38.85 & 38.52 & \best{39.08} & \secondbest{38.93}
        & 39.54 & 39.18 & 39.61 & 39.60 & \secondbest{39.65} & \best{39.66} \\

        &  & {\tiny SSIM$\uparrow$} 
        & 0.945 & 0.942 & 0.950 & 0.949 & \best{0.952} & \secondbest{0.950}
        & 0.954 & 0.950 & 0.953 & 0.953 & \secondbest{0.954} & \best{0.954} \\
        \arrayrulecolor{gray} \cmidrule(lr){2-15}
        
        & \multirow{2}{*}{\tiny $\times 8$} & {\tiny PSNR$\uparrow$} 
        & 35.98 & 35.61 & 36.31 & 35.71 & \best{36.45} & \secondbest{36.34}
        & 36.46 & 36.04 & 36.72 & 36.41 & \secondbest{36.76} & \best{36.87} \\

        &  & {\tiny SSIM$\uparrow$} 
        & 0.920 & 0.914 & \secondbest{0.924} & 0.917 & \best{0.925} & 0.923
        & 0.924 & 0.919 & \secondbest{0.926} & 0.921 & 0.925 & \best{0.929} \\
        
        \arrayrulecolor{black} \midrule

        \multirow{6}{*}{\rotatebox{90}{\parbox{2.03cm}{\centering \textbf{Coronal PD-FS}}}}
        & \multirow{2}{*}{\tiny $\times 4$} & {\tiny PSNR$\uparrow$}   
        & 35.09 & 35.10 & \secondbest{35.31} & 35.23 & 35.23 & \best{35.38}
        & 35.31 & 35.23 & 35.37 & 35.23 & \secondbest{35.42} & \best{35.54} \\

        &  & {\tiny SSIM$\uparrow$} 
        & 0.849 & 0.847 & \best{0.851} & 0.848 & 0.847 & \secondbest{0.851}
        & \best{0.851} & 0.847 & 0.848 & 0.849 & 0.847 & \secondbest{0.849} \\
        
        \arrayrulecolor{gray} \cmidrule(lr){2-15}

        & \multirow{2}{*}{\tiny $\times 6$} & {\tiny PSNR$\uparrow$} 
        & 34.17 & 34.05 & 34.26 & 34.27 & \secondbest{34.29} & \best{34.44} 
        & 34.48 & 34.25 & 34.53 & 34.33 & \secondbest{34.54} & \best{34.59}   \\  
        
        &  & {\tiny SSIM$\uparrow$} 
        & 0.821 & 0.817 & 0.821 & \secondbest{0.824} & 0.822 & \best{0.825} 
        & \secondbest{0.823} & 0.820 & 0.822 & \best{0.823} & 0.822 & 0.822   \\  
        \arrayrulecolor{gray} \cmidrule(lr){2-15}

        & \multirow{2}{*}{\tiny $\times 8$} & {\tiny PSNR$\uparrow$} 
        & 33.11 & 32.86 & 33.21 & 33.06 & \secondbest{33.27} & \best{33.36} 
        & 33.09 & 32.71 & 33.35 & 33.09 & \secondbest{33.48} & \best{33.50}   \\  
        
        &  & {\tiny SSIM$\uparrow$} 
        & 0.794 & 0.791 & 0.795 & 0.796 & \secondbest{0.797} & \best{0.797} 
        & \secondbest{0.796} & 0.785 & \best{0.796} & 0.789 & 0.794 & 0.794   \\  
        
        \arrayrulecolor{black} \midrule

        \multirow{6}{*}{\rotatebox{90}{\parbox{1.8cm}{\centering \textbf{Axial T2}}}}
        & \multirow{2}{*}{\tiny $\times 4$} & {\tiny PSNR$\uparrow$}   
        & 36.37 & 36.42 & \secondbest{36.60} & 36.54 & 36.59 & \best{36.60} 
        & 36.63 & 36.53 & \secondbest{36.81} & 36.75 & 36.77 & \best{36.81} \\
        
        &  & {\tiny SSIM$\uparrow$} 
        & 0.927 & 0.926 & \best{0.928} & 0.928 & 0.925 & \secondbest{0.928}
        & 0.926 & 0.923 & 0.925 & 0.926 & \secondbest{0.926} & \best{0.927}   \\  
        \arrayrulecolor{gray} \cmidrule(lr){2-15}

        & \multirow{2}{*}{\tiny $\times 6$} & {\tiny PSNR$\uparrow$} 
        & 34.53 & 34.69 & \secondbest{35.05} & 34.91 & 35.03 & \best{35.09} 
        & 35.07 & 34.94 & 35.35 & 35.10 & \secondbest{35.37} & \best{35.44} \\
        
        &  & {\tiny SSIM$\uparrow$} 
        & 0.903 & 0.910 & \secondbest{0.910} & \best{0.910} & 0.906 & 0.909
        & \best{0.913} & 0.906 & 0.910 & 0.909 & 0.909 & \secondbest{0.910} \\  
        \arrayrulecolor{gray} \cmidrule(lr){2-15}

        & \multirow{2}{*}{\tiny $\times 8$} & {\tiny PSNR$\uparrow$} 
        & 32.90 & 32.70 & \secondbest{33.41} & 32.98 & 33.26 & \best{33.41} 
        & 33.15 & 32.99 & 33.43 & 33.14 & \best{33.67} & \secondbest{33.56} \\
        
        &  & {\tiny SSIM$\uparrow$} 
        & 0.889 & 0.889 & \best{0.893} & 0.890 & 0.890 & \secondbest{0.892}
        & \best{0.894} & 0.885 & 0.890 & 0.889 & 0.891 & \secondbest{0.891} \\

        \arrayrulecolor{black} \bottomrule
    \end{tabular}
    \vspace{-0.2cm}
\end{table}

\begin{figure}[t]
    \centering
    \includegraphics[width=0.94\linewidth]{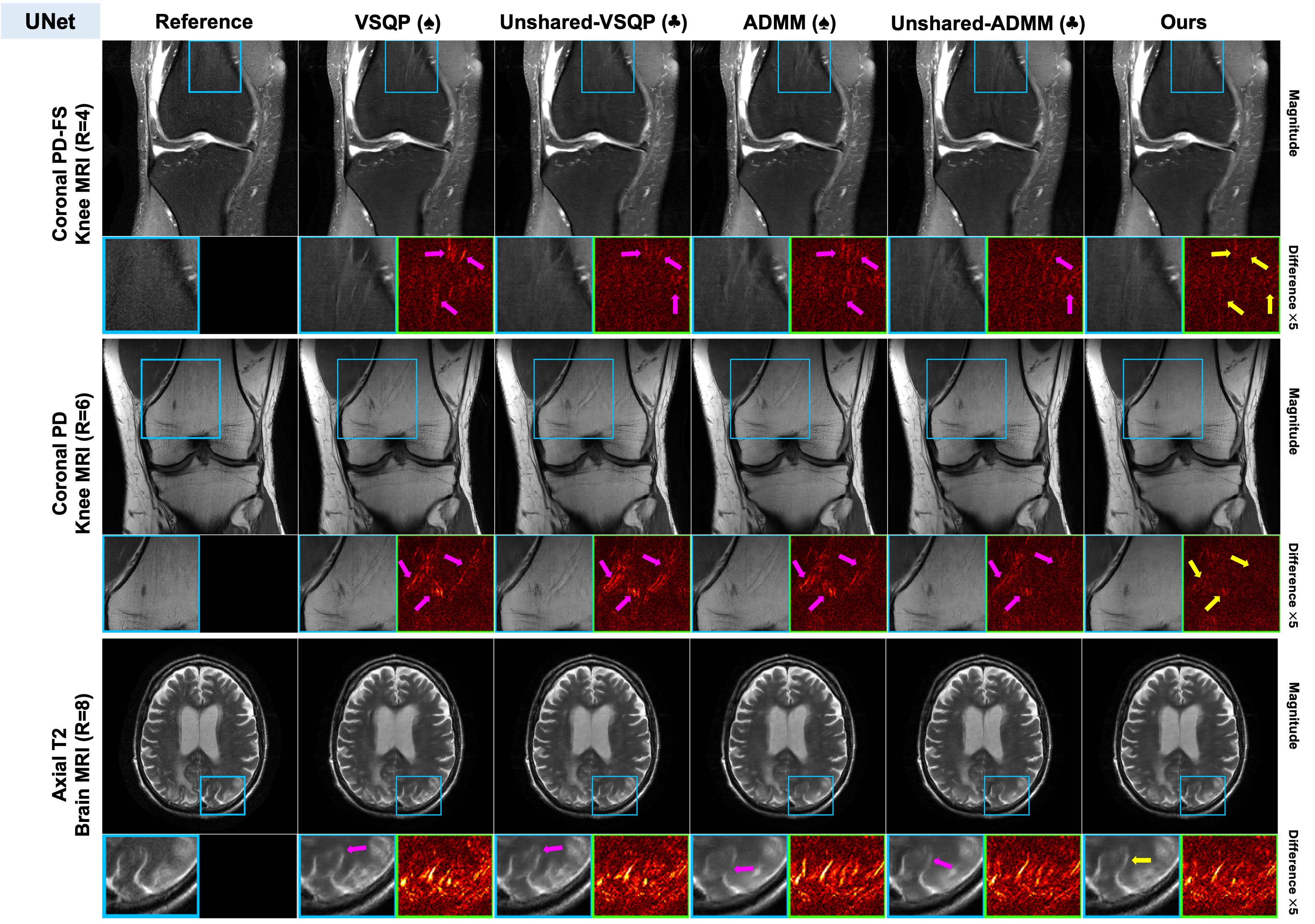} 
    \caption{Qualitative comparisons between the standard shared ($\spadesuit$) and unshared ($\clubsuit$) $\mathcal{R}(\cdot)$ optimization methods (VS, ADMM) and the proposed time-embedded unrolled algorithms with $T\!=\!10$ unrolls in \textbf{U-Net}. \textbf{Top:} Results for $R=4$ using PD-FS data. \textbf{Middle:} Results for $R=6$ using PD data. \textbf{Bottom:} Results for $R=8$ using Axial T2-W data. The proposed methods reduce artifacts (yellow arrow) that the shared and unshared methods fail to eliminate (pink arrow).}
    \vspace{-0.2cm}
    \label{fig:result_unet}
    \vspace{-0.3cm}
\end{figure}

\subsection{Neural Network Architectures for Time-Embedded Proximal Operators} \label{sec:3_2}
The U-Net architecture~\cite{ronneberger2015u} has been widely used, especially in the context of diffusion models, as a time-embedded network by integrating time-embedding features into different layers~\cite{ho2020ddpm,song2021scoreSDE,dhariwal2021beatGANs,ho2021classifierfree}, as shown in \figref{fig:time_emb_networks} (c). We follow the time-embedded U-Net design based on ADM from~\cite{dhariwal2021beatGANs}, which employs group normalization and the FiLM method, as formulated in \eqnref{eq:FiLM-1} and \eqnref{eq:FiLM-2}, with modifications in the number of channels and up/down sampling. We also note that U-Net has connections with message passing through belief propagation ~\cite{mei2025unets}.

We additionally propose a novel time-embedding module for ResNet~\cite{he2016deep}, which is commonly used as a proximal operator in unrolling algorithms for MR reconstruction~\cite{yaman2020SSDU,yaman2022zeroshot}, as shown in \figref{fig:time_emb_networks} (b) and (c). The time-embedding module in ResNet is designed as:
\begin{equation}
    \mathcal{H}_i^t = \mathcal{F}_i^t + \tau \times (\alpha_i^t \odot \text{GroupNorm}(\mathcal{F}_i^t) \oplus \beta_i^t),
\label{eq:FiLM-3}
\end{equation}
where \( \alpha_i^t \) and \( \beta_i^t \) are as in \eqnref{eq:FiLM-1}, and \( \tau \) is a scaling factor. Instead of directly applying the transformed features from \( \alpha_i^t \) and \( \beta_i^t \), this module utilizes \( \tau \) as a scaling factor to indirectly influence the features. This approach ensures a stable integration of time information into the ResNet architecture.

\begin{figure}[t]
    \centering
    \includegraphics[width=0.94\linewidth]{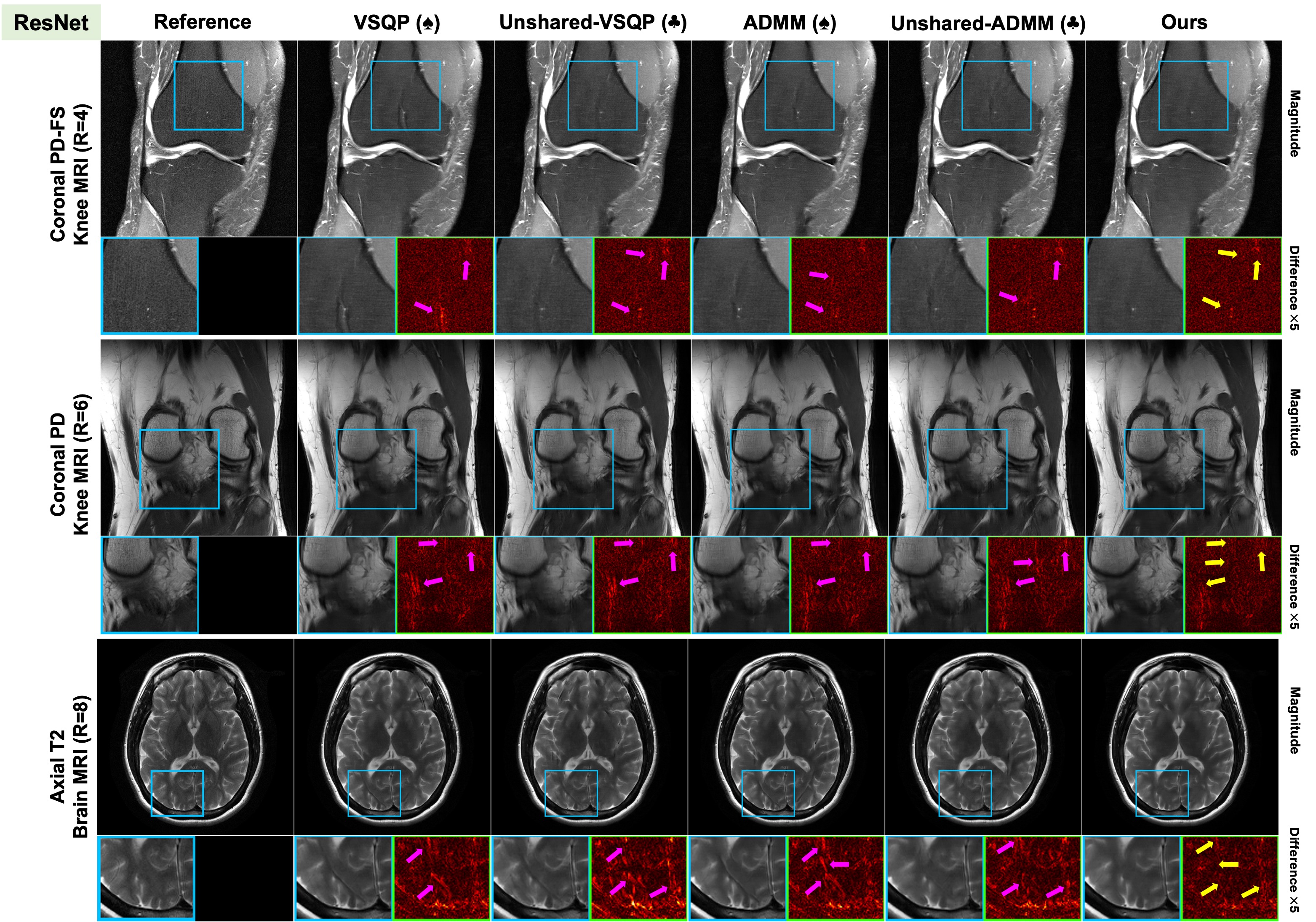} 
    \caption{Qualitative comparisons using ResNet (instead of U-Net in \figref{fig:result_unet}). The proposed methods reduce artifacts (yellow arrow) that the shared and unshared methods fail to eliminate (pink arrow).}
    \vspace{-0.2cm}
    \label{fig:result_resnet}
    \vspace{-0.3cm}
\end{figure}

\section{Experiments and Results} \label{sec:results}
\subsection{Experimental Setup} \label{sec:4_1}
We carried out an in-depth assessment of our approach, analyzing its effectiveness quantitatively and visually through multiple acceleration rates and datasets. The data included fully sampled coronal proton density (PD) and PD with fat-suppression (PD-FS) knee MRI scans, as well as axial T2-weighted brain MRI scans. These scans were obtained from the New York University (NYU) fastMRI database~\cite{knoll2020fastmri_dataset-journal,zbontar2019fastmri_dataset-arXiv}, and were acquired with appropriate institutional review board approvals.

All datasets were retrospectively undersampled using uniform/equidistant undersampling at acceleration factors of $R = 4, 6, \text{ and } 8$ with 24 central kspace lines kept. For this study, we focused on uniform/equidistant undersampling patterns, as they are more commonly used in clinical practice and produce coherent artifacts that are more challenging to remove~\cite{hammernik2023SPM}. We additionally evaluate the generalization performance of our method on random undersampling patterns in~\secref{sec:extended_exp}. For knee datasets, model training was conducted using 300 slices from 10 subjects, while testing was carried out on 380 slices from a separate set of 10 subjects~\cite{hammernik2018VarNet}. For the brain dataset, training and testing were performed using 300 slices each.

\subsection{Implementation Details} \label{sec:4_2}
We compared our method with conventional algorithm unrolling based on VSQP and ADMM, which were implemented with both shared~\cite{aggarwal2019MoDL,yaman2020SSDU} and unshared~\cite{hammernik2018VarNet, muckley2021fastMRIchallenge-2} weights across iterations. We note that multiple variants of ADMM and VSQP unrolling~\cite{yaman2020SSDU, yang2020admm_csnet, yiasemis2025vsharp, duan2019vs-net} have been proposed with different names, primarily differing in their choice of network architectures for the proximal step and their training strategies. In this work, our focus is not on comparing these variations, but rather on analyzing the effect of the outer algorithm unrolling itself with matching proximal operator network structures and training processes. For the least squares problem in the data fidelity of these approaches, conjugate gradient (CG) with 15 iterations was utilized~\cite{aggarwal2019MoDL}. For the proximal operators, we chose two distinct network architectures: 1) a ResNet model with 15 residual blocks, where each block consists of $3\!\times\!3$ convolutional layers with 64 channels~\cite{yaman2020SSDU}, and 2) a U-Net model, adapted from the ADM diffusion model~\cite{dhariwal2021beatGANs} with slight modifications to number of channels and up/down sampling layers.
\begin{wrapfigure}[7]{r}{0.54\textwidth}
\vspace{-2.8em}
\begin{minipage}{0.54\textwidth}
    \begin{table}[H]
    \scriptsize
    \centering
    \setlength{\tabcolsep}{2.2pt}
    \renewcommand{\arraystretch}{0.8}
    \caption{The number of parameters for the shared ($\spadesuit$), unshared ($\clubsuit$), and our proposed methods using different networks with $T\!=\!$ 10 unrolls.}
    \begin{tabular}{@{}c|cc|cc|c@{}}
        \toprule
        \textbf{Networks} & \textbf{VSQP ($\spadesuit$)} & \textbf{VSQP ($\clubsuit$)} & \textbf{ADMM ($\spadesuit$)} & \textbf{ADMM ($\clubsuit$)} & \textbf{Ours} \\ \midrule

        \textbf{U-Net}
        & 1,724,035 & 17,240,341 & 1,724,036 & 17,240,342 & 1,963,479  \\
        \arrayrulecolor{gray} \cmidrule(lr){1-6}

        \textbf{ResNet}
        & 592,129 & 5,921,281 & 592,130 & 5,921,282 & 866,581  \\

        \arrayrulecolor{black} \bottomrule
    \end{tabular}
\label{tab:table_parameters}
\vspace{-0.1cm}
\end{table}

%


\end{minipage}
\end{wrapfigure}
Details about model architectures and hyperparameters are provided in \appref{sec:implementation_details}. The comparisons were first divided by the proximal network architecture, \ie ResNet vs U-Net based. For each of these two proximal network architectures, we trained five unrolled networks from scratch: the proposed time-embedded unrolling, VSQP with shared parameters and unshared parameters, ADMM with shared and unshared parameters. Note that incorporating time embedding into the shared proximal networks leads to a modest increase in parameters, as shown in \tabref{tab:table_parameters}, which is considerably smaller than the increase caused by using a distinct unshared regularizer at each unroll. This shows the efficacy of our proposed approach, which adapts the regularizer over time without significantly increasing network size.

\subsection{Performance of Time-Embedded Unrolling versus Existing Methods} \label{sec:4_3}

\paragraph{Quantitative Results} \tabref{tab:table_quant_main} depicts the performance of different approaches on the Coronal PD, PD-FS, and Axial T2 datasets across different acceleration rates. The shared and unshared baselines are trained with $T\!=\!$ 10 unrolls, while our proposed method is trained with both $T\!=\!$ 5 and 10 unrolls. In almost all cases, the unshared baselines perform worse than their shared counterparts for both U-Net and ResNet in this \emph{limited data} setting. Though unshared methods are known to generalize well in large data regimes~\cite{muckley2021fastMRIchallenge-2}, in our limited data setting, they exhibit performance degradation due to the high number of trainable parameters, as further detailed in \appref{sec:unshared_overfitting} with experiments on fine-tuning from pretrained shared baselines. Our proposed method with $T\!=\!$ 10 unrolls outperforms both shared and unshared methods, achieving the best or second-best performance across all acceleration rates and datasets. The only exceptions are SSIM on Coronal PD at  $R=8$ and Axial T2 at $R=6$ when using the U-Net proximal operator, and SSIM on Coronal PD-FS at $R=6$ and $R=8$ when using the ResNet proximal operator. These results demonstrate that our proposed method performs best in the limited training data regime for a fixed number of unrolls.

Remarkably, even with $T\!=\!$ 5 unrolls, our proposed method achieves performance comparable to the shared baselines with $T\!=\!$ 10 unrolls in most cases. The only notable exception is the Axial T2 dataset, where a small performance gap remains compared to the top-performing methods. This shows that our method can deliver strong performance while halving the number of network computations and reducing inference time by $\sim$50\%, offering a substantial advantage in clinical applications.

 Overall, ResNet-based unrolling networks demonstrate stronger quantitative performance than U-Net-based ones. Additionally, there are a few cases where our performance does not fall within the second-best range. Nevertheless, we note that PSNR and SSIM do not necessarily capture finer details, as noted in earlier studies~\cite{blau2018perception,muckley2021fastMRIchallenge-2, knoll2020fastMRIchallenge-1,chung2023dps,alcalar2024ZAPS}.
Therefore, in the next section, we provide qualitative results to further demonstrate the effectiveness of our approach.

\paragraph{Qualitative Results} \figref{fig:result_unet} and \figref{fig:result_resnet} depict representative reconstructions from different unrolling approaches with $T\!=\!10$ unrolls using U-Net and ResNet based proximal operators, respectively. The shared VSQP and ADMM exhibit artifacts for both proximal operators in various cases. The unshared versions of these methods, which consist of 10 independent regularizers, cannot properly mitigate these artifacts (arrows). In contrast, our proposed method with the time-embedded proximal operators effectively addresses these artifacts across all acceleration rates and datasets. In addition to artifact reduction, our proposed method also enhances image sharpness, most clearly visible in the Axial T2 example in \figref{fig:result_unet}. More visual examples are provided in \appref{sec:additional_results}.

To further investigate the subtle and diagnostically important improvements afforded by our method, representative reconstructions were reviewed by an expert musculoskeletal radiologist, who was blinded to the reconstruction method. The reviewed images included the cases shown in \figref{fig:result_unet} and \figref{fig:result_resnet}, as well as sample annotated pathology cases from the fastMRI+~\cite{zhao2022fastmri_plus} dataset. The radiologist noted that our method was able to remove subtle artifacts that were observed with the other methods in all the reviewed cases.
A detailed description of the readings and the artifacts, as well as the pathological region assessments on fastMRI+ are provided in~\appref{sec:additional_results}. These evaluations further confirm the strong performance of our approach beyond standard quantitative metrics.

\subsection{Extension to Other Unrolling Algorithms} \label{sec:4_4}
\begin{wrapfigure}[11]{r}{0.50\textwidth}
\vspace{-1.6em}
\begin{minipage}{0.50\textwidth}
    \vspace{-0.6em}
    \begin{table}[H]
    \scriptsize
    \centering
    \setlength{\tabcolsep}{4.2pt}
    \renewcommand{\arraystretch}{0.8}
    \caption{Comparison of shared baseline methods and their time-embedded versions (baseline-TE) using the U-Net proximal operator on coronal PD.}
      
    \begin{tabular}{@{}rc|cc|cc@{}}
        \toprule
        \textbf{R} & & \textbf{VSQP} & \textbf{VSQP-TE} & \textbf{ADMM} & \textbf{ADMM-TE}\\ \midrule
        \multirow{2}{*}{$\times 4$} & PSNR$\uparrow$ 
        & 40.50 & \textbf{40.92} & 40.76 & \textbf{40.87} \\
        
          & SSIM$\uparrow$ 
        & 0.962 & \textbf{0.964} & 0.964 & \textbf{0.964} \\
        \arrayrulecolor{gray} \cmidrule(lr){2-6}

         \multirow{2}{*}{$\times 6$} & PSNR$\uparrow$ 
        & 38.12 & \textbf{38.50} & 38.85 & \textbf{38.87} \\
        
          & SSIM$\uparrow$ 
        & 0.945 & \textbf{0.946} & 0.950 & \textbf{0.951} \\
        \arrayrulecolor{gray} \cmidrule(lr){2-6}
        
         \multirow{2}{*}{$\times 8$} & PSNR$\uparrow$ 
        & 35.98 & \textbf{36.15} & 36.31 & \textbf{36.48} \\
        
          & SSIM$\uparrow$ 
        & 0.920 & \textbf{0.921} & 0.924 & \textbf{0.925} \\
        \arrayrulecolor{black} \bottomrule
    \end{tabular}

    \vspace{-0.2cm}
\label{tab:table_time_embedding_extension}
\vspace{-0.2cm}
\end{table}

\end{minipage}
\end{wrapfigure}
Steps 3 and 5 of \algoref{algo:alg_proposed} are analogous to the VSQP updates in \eqnref{eq:vs_x_sol}-\eqref{eq:vs_z_sol}, except with a time-dependent quadratic penalty parameter in the former and a time-dependent proximal operator in the latter. Thus, we set out to explore the effect of the Onsager correction in Step 4 of \algoref{algo:alg_proposed}, specifically to empirically characterize whether the correction to the data fidelity output is minimal, \ie, $\mathbf{x}^{t+1} \approx \mathbf{u}^{t+1}$. Indeed, we observed that the network made only minor differences between $\mathbf{x}^{t+1} $ and $\mathbf{u}^{t+1}$ over iterations. Details are provided in \appref{sec:supp_x_r_comparison}. Omitting the Onsager correction in Step 4 turns \algoref{algo:alg_proposed} to a time-embedded version of VSQP. Similarly, we can also unroll other algorithms, such as ADMM in the proposed time-embedded manner. As shown in \tabref{tab:table_time_embedding_extension}, time-embedded proximal units and data fidelity parameters improve the performance of VSQP and ADMM across all acceleration rates. Like our method in \secref{sec:4_3}, other time-embedded unrolled networks also exhibit superior performance in effectively reducing artifacts. Further details are presented in \appref{sec:supp_wo_w_time_embedding}. 

\subsection{Ablation Studies}  \label{sec:4_5}
We conducted three ablation studies to evaluate the effects of varying hyperparameters of time-embedded unrolled networks. 

\paragraph{Effect of Varying the Numbers of Unrolls}
We trained all the methods in \tabref{tab:table_quant_main} for  $T \in \{5,15\}$. Our proposed method maintained stable performance regardless of the number of unrolls, consistently reducing artifacts, while yielding sharp images, whereas other baselines exhibited varying performance depending on the number of iterations. Further details are given in \appref{sec:supp_detail_ablation}.

\paragraph{Efficiency Analysis with Respect to the Number of Parameters} 
We investigated whether increasing the number of trainable parameters improves performance, in the shared and unshared baselines. Increasing parameters do not improve results in ResNet, whereas a slight quantitative gain is observed for U-Net, though visual residual artifacts persist. These larger models also incur higher computational costs, whereas our method achieves better performance with only a marginal increase in parameters. Detailed experimental results are provided in ~\appref{sec:supp_detail_ablation} \tabref{tab:ablation_number_param}.

\paragraph{Time-Embedding Module with Different Hyperparameters}
We conducted experiments with a time-embedded U-Net to evaluate key hyperparameters of the time-embedding module, including the sinusoidal encoding frequency, embedding dimension, and the number of hidden channels in the MLP layers. Performance was influenced by these hyperparameters, with optimal results achieved using a period of 10,000, an embedding dimension of 32, and 128 hidden channels. These settings were applied consistently across all experiments. Additional information is in~\appref{sec:supp_detail_ablation} \tabref{tab:time_emb_hyperparameter}.

\subsection{Extended Experiments} \label{sec:extended_exp}

\paragraph{Artifact Evolution Across Unrolls}

While the time step $t$ in diffusion models explicitly determines the noise level at each stage of the forward process, the time step $t$ in our setting plays a different role. $t$ implicitly governs the evolution of the proximal operator across iterations by accounting for the changing distribution of intermediate features, analogous to the Onsager correction term in VAMP, which stabilizes updates by compensating for iterative correlations. This enables time-embedded proximal operators with temporal information to adaptively apply varying levels of denoising at different stages, which is further illustrated in \appref{sec:supp_wo_w_time_embedding} \figref{fig:appdx_stage_wise_results}.

\paragraph{Validation on Non-Uniform Sampling Masks} 
To evaluate the effectiveness of our proposed method under non-uniform (random) sampling patterns, we conducted experiments at various acceleration rates, using both baseline methods and our approach with a U-Net architecture and 10 unrolls. These results confirm that the effectiveness of our method extends to non-uniform undersampling patterns. Results are provided in~\appref{sec:supp_extended_exp}.

\paragraph{Comparison with Diffusion-Based Models}
Since diffusion-based reconstruction provides a promising approach for solving MR inverse problems~\cite{jalal2021robust, chung2024decomposed, alccalar2025ZADS, gulle2025consistency}, we compared our results with Decomposed Diffusion Sampling (DDS)~\cite{chung2024decomposed}. Our method outperforms DDS in terms of both PSNR and SSIM. Details of the implementation and the corresponding results are provided in~\appref{sec:supp_extended_exp}.

\section{Limitations and Discussion} \label{sec:limit_discuss}
\textbf{Limitations.} 
As discussed in \secref{sec:intro} and \secref{sec:results}, our experiments were conducted in a limited data regime, using 300 slices per dataset. This setting is particularly relevant for translational applications, where new imaging sequences are being developed or when higher resolutions are targeted. While our method demonstrated strong performance and generalization in this regime, the performance gap between our approach and unshared baselines may narrow when training with more data samples, as the risk of overfitting will be lower for the latter.

\textbf{Discussion.} 
Through empirical evaluation, we examined whether second-moment matching holds in our time-embedded unrolling algorithms inspired by the Vector AMP framework, as well as the Lipschitz constants and stability of the time-embedded FiLM layers. Detailed discussions are provided in~\appref{sec:discuss}.

\section{Conclusion}

In this study, we introduced a time-embedded algorithm unrolling framework inspired by AMP theory and time-embedding in diffusion models. Our unrolled networks used time-embedding in proximal operators, which performed both denoising and Onsager correction, as well as in data fidelity weights. We extended these ideas to VSQP and ADMM-based unrolling, demonstrating the framework's versatility. Our method outperformed both shared and unshared unrolling approaches under matched settings, producing sharper images with fewer artifacts, especially in limited data regime. Unlike unshared models, which showed signs of overfitting, our method generalized better and remained robust across different unroll depths.

\section{Acknowledgements}
The authors gratefully acknowledge Dr. Jutta Ellermann for providing expert radiologist assessments during the rebuttal stage on short notice.
This work was partially supported by NIH R01HL153146, NIH R01EB032830, NIH P41EB027061.

\bibliographystyle{abbrv}
\small
\bibliography{refs_utku_base}

\newpage
\section*{NeurIPS Paper Checklist}

\begin{enumerate}

\item {\bf Claims}
    \item[] Question: Do the main claims made in the abstract and introduction accurately reflect the paper's contributions and scope?
    \item[] Answer: \answerYes{} 
    \item[] Justification: The claims we made in abstract and introduction accurately reflect the paper’s
contributions and scope.

\item {\bf Limitations}
    \item[] Question: Does the paper discuss the limitations of the work performed by the authors?
    \item[] Answer: \answerYes{} 
    \item[] Justification: We have provided a detailed discussion of our study’s limitations in Appendix \secref{sec:limit_discuss}.

\item {\bf Theory assumptions and proofs}
    \item[] Question: For each theoretical result, does the paper provide the full set of assumptions and a complete (and correct) proof?
    \item[] Answer: \answerYes{} 
    \item[] Justification:  All assumptions and proofs can be found in \secref{sec:Methodology}.

    \item {\bf Experimental result reproducibility}
    \item[] Question: Does the paper fully disclose all the information needed to reproduce the main experimental results of the paper to the extent that it affects the main claims and/or conclusions of the paper (regardless of whether the code and data are provided or not)?
    \item[] Answer: \answerYes{} 
    \item[] Justification: We have provided a full description of the algorithms used in the paper, with detailed explanations included in \secref{sec:results} and the Appendix \secref{sec:implementation_details}.

\item {\bf Open access to data and code}
    \item[] Question: Does the paper provide open access to the data and code, with sufficient instructions to faithfully reproduce the main experimental results, as described in supplemental material?
    \item[] Answer: \answerYes{} 
    \item[] Justification: Data for our retrospective studies are openly available and details to reproduce the main experimental results are provided in \secref{sec:4_1}, \secref{sec:4_2} and Appendix \secref{sec:implementation_details}. The code is available publicly at~\url{https://github.com/JN-Yun/TE-Unrolling-MRI}. 
    

\item {\bf Experimental setting/details}
    \item[] Question: Does the paper specify all the training and test details (e.g., data splits, hyperparameters, how they were chosen, type of optimizer, etc.) necessary to understand the results?
    \item[] Answer: \answerYes{} 
    \item[] Justification: The experimental details are provided in the Appendix \secref{sec:implementation_details}.

\item {\bf Experiment statistical significance}
    \item[] Question: Does the paper report error bars suitably and correctly defined or other appropriate information about the statistical significance of the experiments?
    \item[] Answer: \answerYes{} 
    \item[] Justification: Error bars/standard deviations for the quantitative metrics are reported in the Appendix \secref{sec:std}. 

\item {\bf Experiments compute resources}
    \item[] Question: For each experiment, does the paper provide sufficient information on the computer resources (type of compute workers, memory, time of execution) needed to reproduce the experiments?
    \item[] Answer: \answerYes{} 
    \item[] Justification:  All computational resources are reported in Appendix \secref{sec:implementation_details}.
    
\item {\bf Code of ethics}
    \item[] Question: Does the research conducted in the paper conform, in every respect, with the NeurIPS Code of Ethics \url{https://neurips.cc/public/EthicsGuidelines}?
    \item[] Answer: \answerYes{} 
    \item[] Justification: The research presented in this paper fully complies with the NeurIPS Code of Ethics in all respects.

\item {\bf Broader impacts}
    \item[] Question: Does the paper discuss both potential positive societal impacts and negative societal impacts of the work performed?
    \item[] Answer: \answerYes{} 
    \item[] Justification: We have discussed the potential positive societal impacts of this work in \secref{sec:intro} and \secref{sec:results}. We believe our work does not pose any negative societal impacts. 
    
\item {\bf Safeguards}
    \item[] Question: Does the paper describe safeguards that have been put in place for responsible release of data or models that have a high risk for misuse (e.g., pretrained language models, image generators, or scraped datasets)?
    \item[] Answer: \answerNA{} 
    \item[] Justification: The paper poses no such risks.

\item {\bf Licenses for existing assets}
    \item[] Question: Are the creators or original owners of assets (e.g., code, data, models), used in the paper, properly credited and are the license and terms of use explicitly mentioned and properly respected?
    \item[] Answer: \answerYes{} 
    \item[] Justification: All codes and datasets used in this paper have been properly cited.

\item {\bf New assets}
    \item[] Question: Are new assets introduced in the paper well documented and is the documentation provided alongside the assets?
    \item[] Answer: \answerNA{} 
    \item[] Justification: This paper does not release new assets.

\item {\bf Crowdsourcing and research with human subjects}
    \item[] Question: For crowdsourcing experiments and research with human subjects, does the paper include the full text of instructions given to participants and screenshots, if applicable, as well as details about compensation (if any)? 
    \item[] Answer: \answerNA{} 
    \item[] Justification: This paper does not involve crowdsourcing nor research with human subject.

\item {\bf Institutional review board (IRB) approvals or equivalent for research with human subjects}
    \item[] Question: Does the paper describe potential risks incurred by study participants, whether such risks were disclosed to the subjects, and whether Institutional Review Board (IRB) approvals (or an equivalent approval/review based on the requirements of your country or institution) were obtained?
    \item[] Answer: \answerYes{} 
    \item[] Justification: Knee and brain imaging data was obtained from the publicly available NYU fastMRI dataset~\cite{knoll2020fastmri_dataset-journal,zbontar2019fastmri_dataset-arXiv}, which was collected with IRB approval and subject consent as detailed in the original publication. No additional data involving human subjects were collected in this study.

\item {\bf Declaration of LLM usage}
    \item[] Question: Does the paper describe the usage of LLMs if it is an important, original, or non-standard component of the core methods in this research? Note that if the LLM is used only for writing, editing, or formatting purposes and does not impact the core methodology, scientific rigorousness, or originality of the research, declaration is not required.
   
    \item[] Answer: \answerNA{} 
    \item[] Justification: This work does not involve the use of LLMs in any core method or experimental component.

\end{enumerate}

\newpage
\appendix
{\Large \textbf{Appendix}}\par

\section{Model Architectures and Relevant Hyperparameters} \label{sec:implementation_details}

\paragraph{Neural Network Architectures.} A ResNet and a U-Net architecture was used for proximal operators in the unrolled networks, as illustrated in \figref{fig:time_emb_networks} (c). The ResNet used for the proximal operator \cite{timofte2017ntire} consists of 15 residual blocks, each containing \(3 \!\times\! 3\) convolutional layers with ReLU activation and a scaling term. The scaling term is set to \(1 \!\times\! 10^{-1}\) \cite{yaman2020SSDU}. The U-Net used for the proximal operator, which is designed based on ADM from~\cite{dhariwal2021beatGANs} with slight modifications, has 2 downsampling layers, 2 upsampling layers, and a bottleneck. It uses residual blocks with \(3 \!\times\! 3\) convolutional layers, normalization, and SiLU activation. The initial channel size is 32, which doubles during downsampling and is recovered during upsampling. For time-embedded architectures, time-embedded features are injected and modulated through group normalization and the FiLM method in each residual block, as shown in \figref{fig:time_emb_networks} (b) and (c). All training processes are conducted using one NVIDIA A100-SXM4-40GB GPU.

\paragraph{Shared/Unshared Baseline.} For the data fidelity term, we use a shared \( \mu \) initialized to \(5\!\times\!10^{-2}\) in VSQP and \(1.5\!\times\!10^{-2}\) in ADMM. In ADMM, we also initialized the dual update parameter \( \lambda \) to \(1\!\times\!10^{-1}\). For the shared baseline networks, both a ResNet and U-Net proximal operator as described above were used, without time-embedding features. In the unshared case, there is a separate proximal operator for each unroll with no weight sharing or time-embedding. We train ResNet for 100 epochs and U-Net for 50 epochs, using a learning rate of \(5\! \times \! 10^{-4}\) for coronal PD/PD-FS knee data and \(2 \! \times 10^{-4} \! \) for Axial T2 brain data with the Adam optimizer. 

\begin{table}[!b]
    \caption{Comparison of the results from the fine-tuned \textbf{Unshared} (ADMM) method with those of shared and unshared baselines trained from scratch in \textit{limited data} settings. \textbf{FS}: From Scratch; \textbf{FT}: Fine-Tuning. Quantitative results are reported across three datasets with varying undersampling patterns. The best values are highlighted in \textbf{bold}.}
    \label{tab:table_unshared_overfitting}
    \centering
    \scriptsize
    \setlength{\tabcolsep}{2.0pt}
    \renewcommand{\arraystretch}{0.70}
    \begin{tabular}{@{}p{0.2cm}rcc|c|ccccc|c|c|ccccc@{}}
        \toprule
         &  &  & \multicolumn{7}{c}{{\bf U-Net}} & \multicolumn{7}{c}{{\bf ResNet}}\\
         \cmidrule(lr){4-10} \cmidrule(lr){11-17}
         &  & & {\bf FS \tiny (Shared)} & {\bf FS \tiny (Unshared)} & \multicolumn{5}{c}{{\bf FT \tiny (Unshared)}} & {\bf FS \tiny (Shared)} & {\bf FS \tiny (Unshared)} & \multicolumn{5}{c}{{\bf FT \tiny (Unshared)}}\\
         \cmidrule(lr){4-10} \cmidrule(lr){11-17}
        \textbf{ } &  & \textbf{Epoch} & {\textbf{100}} & {\textbf{100}} & {\textbf{10}} & {\textbf{20}} & {\textbf{30}} & {\textbf{40}} & {\textbf{50}} & {\textbf{100}} & {\textbf{100}} & {\textbf{10}} & {\textbf{20}} & {\textbf{30}} & {\textbf{40}} & {\textbf{50}} \\ \midrule
        \multirow{6}{*}{\rotatebox{90}{\parbox{1.5cm}{\centering \textbf{Coronal PD}}}}
        & \multirow{2}{*}{$\times 4$} & PSNR$\uparrow$ 
        & 40.76 & 40.51 & \bf{40.96} & 40.89 & 40.82 &  40.78 & 40.69
        & 41.27 & 41.11 & \bf{41.45} &  41.37 &  41.28 &  41.25  &  41.18 \\

        &  & SSIM$\uparrow$ 
        & 0.964 & 0.963 & \bf{0.964} &  0.964 &  0.964 &  0.963 & 0.963
        & \bf{0.965} &  0.964 & 0.965 &  0.965 &  0.965 &  0.964 &  0.964 \\
        
        \arrayrulecolor{gray} \cmidrule(lr){2-17}
        
        & \multirow{2}{*}{$\times 6$} & PSNR$\uparrow$ 
        & 38.85 &  38.52 & \bf{39.13} & 39.13 & 39.09 &  38.99 & 38.94
        & 39.61 &  39.60  & \bf{39.88} &  39.78 &  39.72 &  39.72  &  39.64 \\

        &  & SSIM$\uparrow$ 
        & 0.950 & 0.949 & \bf{0.952} &  0.952 &  0.952 &  0.951 & 0.951
        & 0.953 & 0.953 & \bf{0.955} &  0.955 &  0.955 &  0.955 &  0.953 \\
        \arrayrulecolor{gray} \cmidrule(lr){2-17}
        
        & \multirow{2}{*}{$\times 8$} & PSNR$\uparrow$ 
        & 36.31 & 35.71 & \bf{36.51} & 36.29 & 36.29 &  36.25 & 36.11
        & 36.72 & 36.41 & \bf{36.97} &  36.93 &  36.76 &  39.74  &  36.63 \\

        &  & SSIM$\uparrow$ 
        & 0.924 & 0.917 &  \bf{0.925} &  0.923 &  0.923 &  0.922 & 0.920
        & 0.926 & 0.921 & \bf{0.927} &  0.927 &  0.925 &  0.925 &  0.923 \\
        
        \arrayrulecolor{black} \midrule

        \multirow{6}{*}{\rotatebox{90}{\parbox{1.6cm}{\centering \textbf{Coronal PD-FS}}}}
        & \multirow{2}{*}{$\times 4$} & PSNR$\uparrow$   
        & 35.31 & 35.23 & \bf{35.44} &  35.32 &  35.22 &  35.12 &  35.01
        & 35.37 & 35.23 & \bf{35.59} & 35.57 & 35.54 & 35.51 & 35.50 \\

        &  & SSIM$\uparrow$ 
        & \bf{0.851} & 0.848 & 0.850 &  0.849 & 0.847 &  0.844 &  0.841
        & 0.848 & 0.849 & 0.850 & 0.851 & \bf{0.852} & 0.852 & 0.852 \\
        
        \arrayrulecolor{gray} \cmidrule(lr){2-17}

        & \multirow{2}{*}{$\times 6$} & PSNR$\uparrow$ 
        & 34.26 & 34.27 & \bf{34.45} & 34.33 & 34.26 & 34.17 & 34.11
        & 34.53 & 34.33 & \bf{34.69} & 34.67 & 34.64 & 34.61 & 34.58 \\  
        
        &  & SSIM$\uparrow$ 
        & 0.821 & \bf{0.824} & 0.822 & 0.822 & 0.820 & 0.819 & 0.819
        & 0.822 & 0.823 & 0.823 & 0.823 & 0.825 & 0.825 & \bf{0.825}  \\  
        \arrayrulecolor{gray} \cmidrule(lr){2-17}

        & \multirow{2}{*}{$\times 8$} & PSNR$\uparrow$ 
        & 33.21 & 33.06 & \bf{33.34} & 33.18 & 33.05 & 32.96 & 32.92
        & 33.35 & 33.09 & \bf{33.61} & 33.57 & 33.53 & 33.44 & 33.38  \\  
        
        &  & SSIM$\uparrow$ 
        & 0.795 & \bf{0.796} & 0.795 & 0.792 & 0.790 & 0.788 & 0.786 
        & 0.796 & 0.789 & \bf{0.797} & 0.767 & 0.797 & 0.795 & 0.795  \\  
        
        \arrayrulecolor{black} \midrule

        \multirow{6}{*}{\rotatebox{90}{\parbox{1.6cm}{\centering \textbf{Axial T2}}}}
        & \multirow{2}{*}{$\times 4$} & PSNR$\uparrow$   
        & 36.60 &  36.54 & \bf{36.67} & 36.67 & 36.60 & 36.67 & 36.62
        & 36.81 & 36.75 & \bf{36.87} & 36.83 & 36.86 & 36.83 & 36.85 \\
        
        &  & SSIM$\uparrow$ 
        & 0.928 &  \bf{0.928} & 0.927 & 0.927 & 0.927 & 0.927 & 0.927
        & 0.925 & \bf{0.926} & 0.924 & 0.925 & 0.925 & 0.925 & 0.925 \\
        \arrayrulecolor{gray} \cmidrule(lr){2-17}

        & \multirow{2}{*}{$\times 6$} & PSNR$\uparrow$ 
        & 35.05 &  34.91 & \bf{35.16} & 35.16 & 35.15 & 35.06 & 35.10
        & 35.35 & 35.10 & 35.41 & \bf{35.41} & 35.47 & 35.44 & 35.33 \\
        
        &  & SSIM$\uparrow$ 
        & 0.910 & 0.910 & 0.910 & 0.911 & 0.910 & \bf{0.911} & 0.910
        & \bf{0.910} & 0.909 & 0.907 & 0.908 & 0.909 & 0.908 & 0.907 \\
        \arrayrulecolor{gray} \cmidrule(lr){2-17}

        & \multirow{2}{*}{$\times 8$} & PSNR$\uparrow$ 
        & 33.41 & 32.98 & \bf{33.52} & 33.40 & 33.36 & 33.39 & 33.42
        & 33.43 & 33.14 & 33.91 & \bf{33.93} & 33.90 & 33.83 & 33.78 \\
        
        &  & SSIM$\uparrow$ 
        & \bf{0.893} & 0.890 & 0.893 & 0.892 & 0.892 & 0.892 & 0.892
        & 0.890 & 0.889 & 0.890 & 0.890 & 0.890 & 0.890 & \bf{0.890} \\
        
        \arrayrulecolor{black} \bottomrule
    \end{tabular}
\end{table}

\paragraph{Time Embedded Unrolled Networks.} For our proposed time-embedded unrolled networks, as described in \secref{sec:3_1} and \secref{sec:3_2}, we utilized time-dependent data fidelity, Onsager correction parameters, and time-embedded proximal operators. In particular, the data fidelity scalars \( \mu^t \) were initialized to \( 1.5\!\times\!10^{-2} \), and the time-dependent Onsager correction parameters \( \rho^t \) to \( 1\!\times\!10^{-1} \). For time-embedded neural networks, we apply the same optimization strategies as in the baseline, except for ResNet in the coronal PD-FS case, where we use a learning rate of \( 1.8\!\times\!10^{-2} \). The scaling factor $\tau$ of FiLM in ResNet is set to 0.1. To extend our approach to other unrolling algorithms described in \secref{sec:4_4}, we replace the shared data fidelity term \( \mu \) with the unshared data fidelity term \( \mu^t \) and utilize the same proximal operators as in our proposed methods.

\paragraph{Generalization to Diverse Datasets}
Our method incorporates a time-embedding module, which introduces additional hyperparameters. These include the frequency of the sinusoidal encoding, the embedding dimension, and the number of hidden channels in the MLP layers that process the time embeddings. We use the same configuration across different datasets (Coronal PD, Coronal PD-FS, and Axial T2 brain), demonstrating the robustness of our approach to diverse data. Similarly, we apply identical hyperparameters to the neural networks used as proximal operators across all datasets, on which the networks consistently perform well.

\section{Fine-tuned Unshared Methods in Limited Data Settings} \label{sec:unshared_overfitting}

\begin{figure*}
    \centering
    \includegraphics[width=0.9\linewidth]{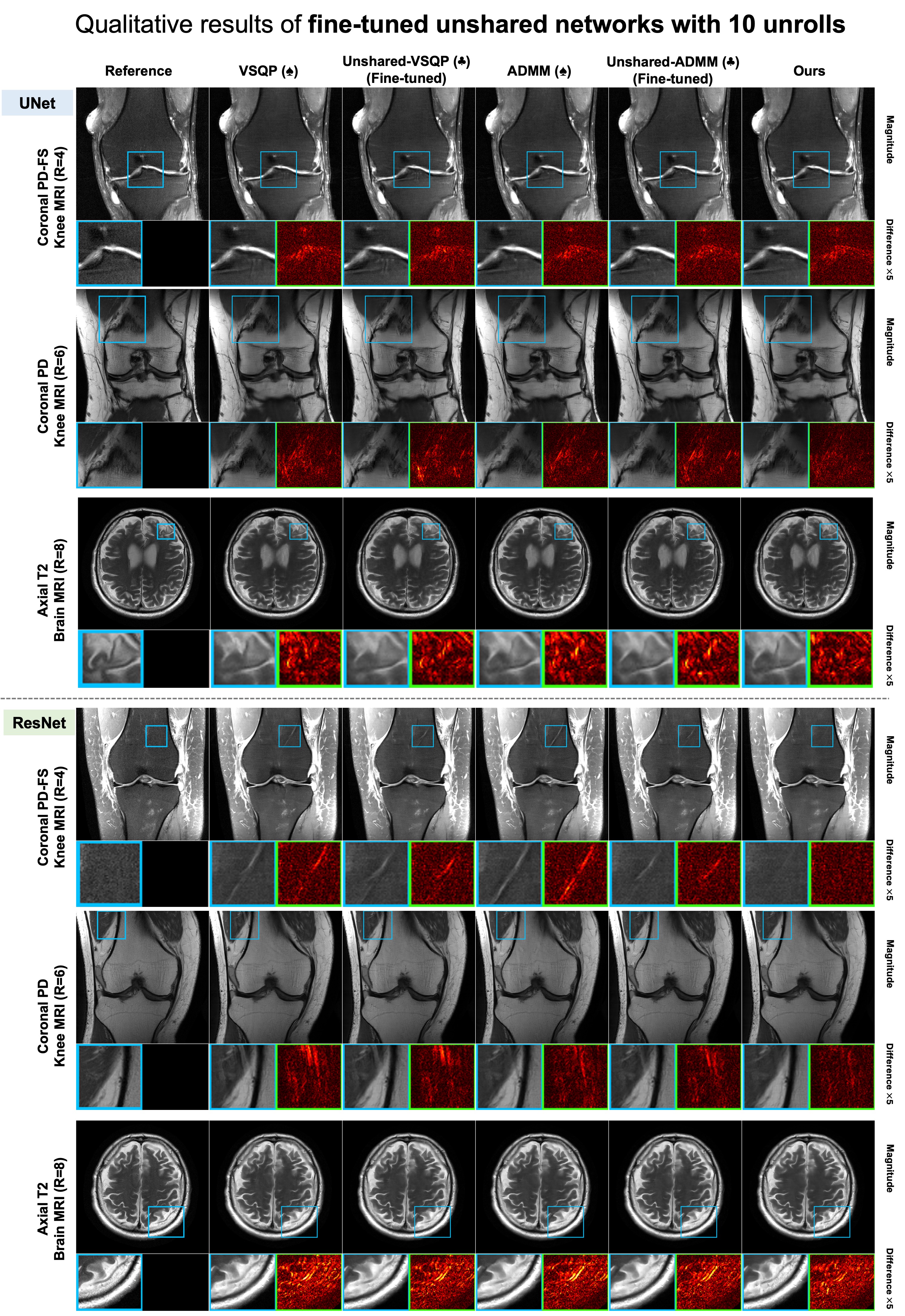} 
    \caption{$\spadesuit$: Shared $\mathcal{R}(\cdot)$ weights, $\clubsuit$: \textbf{Fine-tuned} Unshared $\mathcal{R}(\cdot)$ weights. Qualitative comparisons of each unrolled network with \textbf{$T\!=\!$ 10 unrolls} for \textbf{U-Net} and \textbf{ResNet} proximal operators. In each proximal operator, \textbf{Top:} Results for $R=4$ using PD data. \textbf{Middle:} Results for $R=6$ using PD-FS data. \textbf{Bottom:} Results for $R=8$ using Axial T2-W data. The fine-tuned unshared networks still struggle to suppress artifacts over iterations, whereas the proposed methods perform well, effectively reducing artifacts.}
    \label{fig:appdx_fine_tuned_unshared} 
\end{figure*}

To mitigate the overfitting tendency of unshared networks in limited data settings, we explore fine-tuning instead of training from scratch. Specifically, We initialize the unshared baseline unrolled network from the pre-trained shared baseline unrolled network. This unshared unrolled network is then fine-tuned for several epochs with a learning rate of \(1 \times 10^{-4}\) for both ResNet and U-Net, and for knee and brain data comprising 300 slices each. 
As shown in \tabref{tab:table_unshared_overfitting}, the fine-tuned unshared methods generally outperform both shared and unshared methods trained from scratch. However, overfitting remains evident in the fine-tuned models as the number of training epochs increases under limited data conditions. Furthermore, although the fine-tuned unshared methods improve PSNR and SSIM scores, they still struggle to suppress artifacts over iterations (see \figref{fig:appdx_fine_tuned_unshared} for visual examples). 

\section{Analysis of the Onsager Correction Term in Algorithm 1} 
\label{sec:supp_x_r_comparison}

\begin{figure*}[h]
    \centering
    \includegraphics[width=0.98\linewidth]{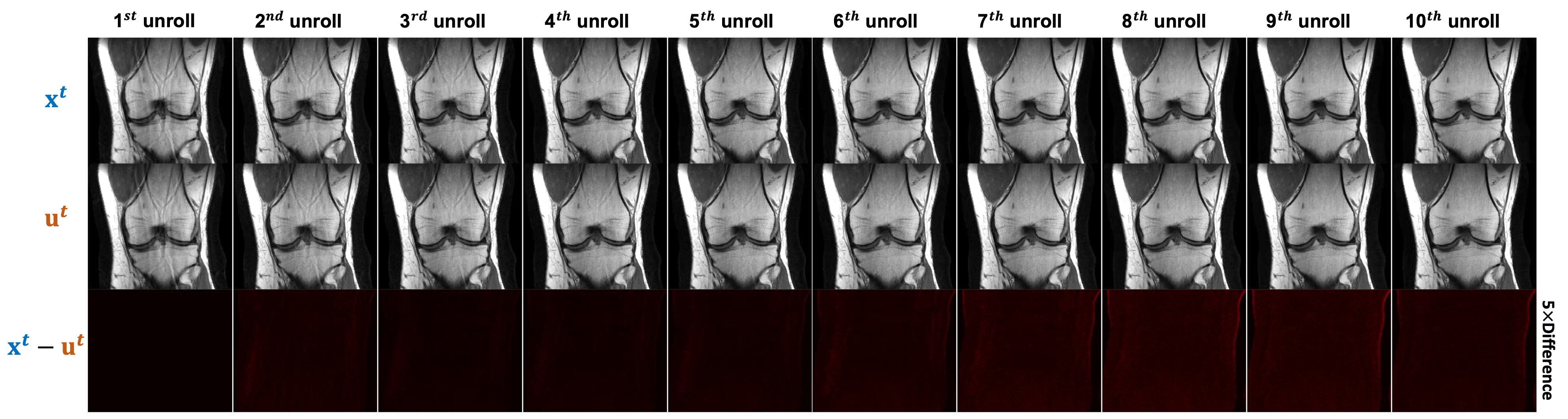} 
    \caption{The intermediate visual results of the proposed method with ResNet-TE proximal operator at each iteration in a coronal PD slice.}
    \label{fig:x_r_comparison}    
\end{figure*}

\begin{table*}[h]
    \fontsize{5}{6}\selectfont
    \caption{Average normalized mean squared error between \( \mathbf{x}^{t} \) and \( \mathbf{u}^{t} \) at each iteration of the unrolled network with ResNet-TE on the coronal PD test set (\( R = 4 \)).}
    \centering
    \setlength{\tabcolsep}{0.8pt}
    \begin{tabular}{@{}ccccccccccccc@{}}
        \toprule
        \textbf{Iteration} & \textbf{1} & \textbf{2} & \textbf{3} & \textbf{4} & \textbf{5} & \textbf{6} & \textbf{7} & \textbf{8} & \textbf{9} & \textbf{10} \\ \midrule
        
        \makecell{\textbf{Nomalized MSE}} 
          
        & $1.01\!\times\!10^{-9}$ 
        & $5.46\!\times\!10^{-3}$ 
        & $9.02\!\times\!10^{-3}$ 
        & $3.77\!\times\!10^{-3}$ 
        & $6.75\!\times\!10^{-3}$ 
        & $1.26\!\times\!10^{-2}$ 
        & $2.55\!\times\!10^{-2}$ 
        & $3.35\!\times\!10^{-2}$ 
        & $2.68\!\times\!10^{-2}$ 
        & $9.96\!\times\!10^{-3}$ \\ 
        
        \arrayrulecolor{black} \bottomrule
    \end{tabular}

\label{tab:table_x_u_comparison}
\end{table*}


To explore the effect of the Onsager correction in Step 4 of Algorithm 1, we evaluated whether the intermediate updates in the network satisfy \( \mathbf{x}^{t} \approx \mathbf{u}^{t} \). We compared the outputs of the data fidelity, (\( \mathbf{x}^{t} \)) and its Onsager correction term output, (\( \mathbf{u}^{t} \)) across unrolls, which is shown in \figref{fig:x_r_comparison}. The bottom row shows the scaled (\(\times 5\)) difference between them, which is minimal upon visual inspection. We further quantified this difference by calculating the normalized mean squared error between \( \mathbf{x}^t \) and \( \mathbf{u}^t \) at each iteration. \tabref{tab:table_x_u_comparison} shows that the difference ranges from \( 1.01\! \times\! 10^{-9} \) to \( 3.35\! \times\! 10^{-2} \), indicating no substantial variation.

\begin{figure*}[t]
    \centering
    \includegraphics[width=0.85\linewidth]{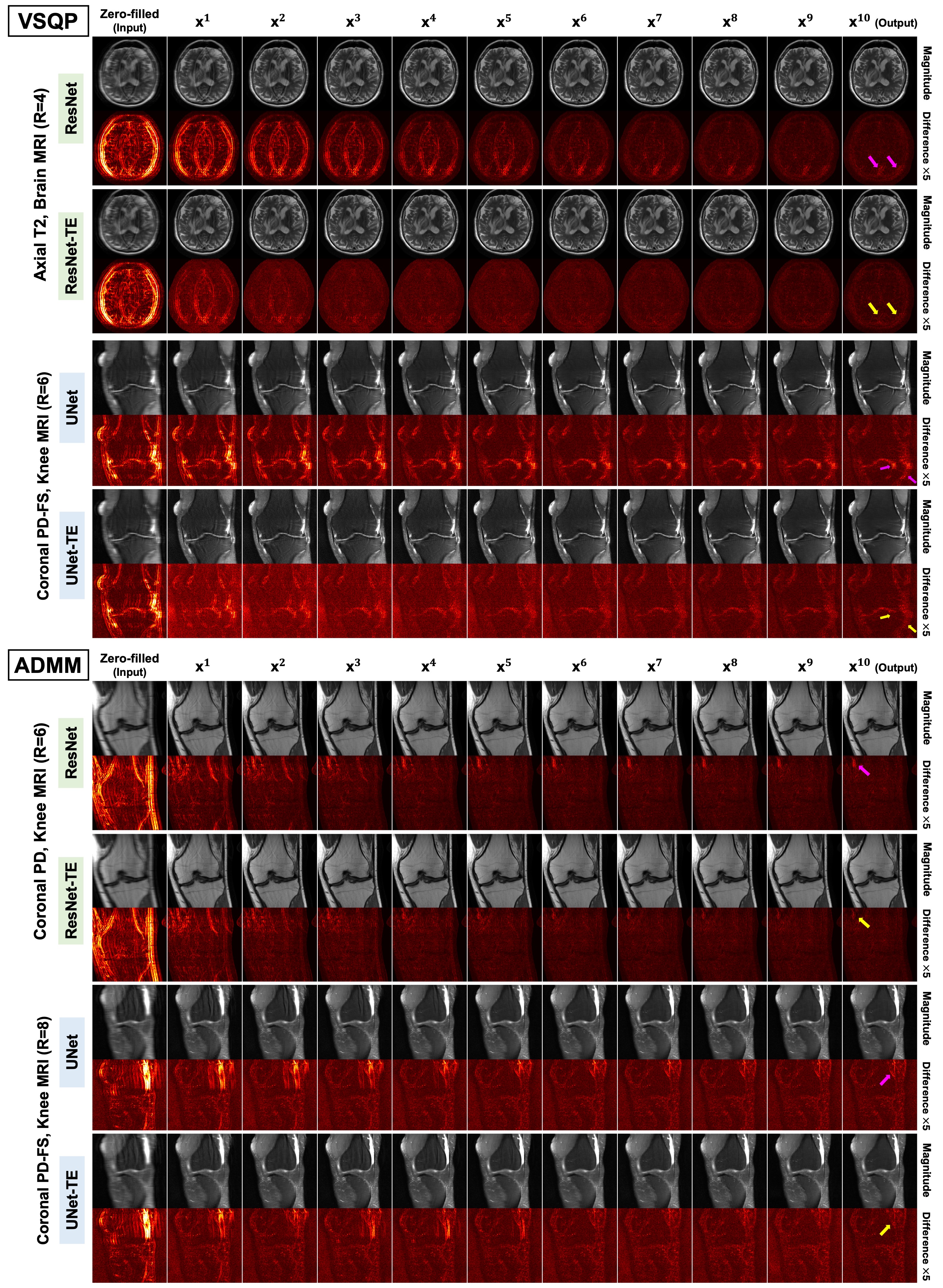} 
    \caption{The estimates through different stages of the unrolled network with shared baseline VSQP, ADMM, and the time-embedded VSQP, ADMM for different datasets with varying acceleration rates. The error maps illustrate the differences between the intermediate estimates, $\{\mathbf{x}^t\}_{t=1}^T$, and the reference. The shared baselines, utilizing both U-Net and ResNet proximal operators, exhibit persistent errors across unrolls, which are highlighted with pink arrows in the last unroll, and can be visualized at the same location in prior unrolls. In contrast, the time-embedded networks, with proximal operators U-Net-TE and ResNet-TE, effectively reduce noise (yellow arrows).}
    \label{fig:appdx_stage_wise_results}
\end{figure*}

\section{Qualitative Comparison of the Baseline and Time-Embedded Algorithm Unrolling} \label{sec:supp_wo_w_time_embedding}
As discussed in \secref{sec:4_4} and in view of \secref{sec:supp_x_r_comparison}, our time-embedding approach can be extended to other unrolled algorithms (VSQP and ADMM) by incorporating a time-embedding module into the proximal operators of these unrolled networks.  The time-embedded versions of the unrolled algorithms are given below. For time-embedded VSQP:
\begin{align}
    \mathbf{x}^{t} &= \left( \mathbf{E}^H_{\Omega} \mathbf{E}_{\Omega} + \mu^t \mathbf{I} \right)^{-1} \left( \mathbf{E}^H_{\Omega} \mathbf{y}_{\Omega} + \mu^t \mathbf{z}^{t}  \right), \label{eq:vs_x_sol_te} \\
    \mathbf{z}^{t+1} &= \text{prox}_\mathcal{R}(\mathbf{x}^{t}, \alpha^t, \beta^t, t), \label{eq:vs_z_sol_te} \\
    \intertext{For time-embedded ADMM:}
    \mathbf{x}^{t+1} &= \left( \mathbf{E}^H_{\Omega} \mathbf{E}_{\Omega} + \mu^t \mathbf{I} \right)^{-1} \left( \mathbf{E}^H_{\Omega} \mathbf{y}_{\Omega} + \mu^t \left( \mathbf{z}^{t} - \mathbf{u}^t \right) \right), \label{eq:admm_x_sol_te} \\
    \mathbf{z}^{t+1} &= \text{prox}_\mathcal{R}(\mathbf{x}^{t+1} + \mathbf{u}^t, \alpha^t, \beta^t, t), \label{eq:admm_z_sol_te} \\
    \mathbf{u}^{t+1} &= \mathbf{u}^{t} + \lambda(\mathbf{x}^{t+1} - \mathbf{z}^{t+1}) ,\label{eq:admm_u_update_te}
\end{align}
where the data fidelity parameter $\mu$ and the proximity operator $\text{prox}_\mathcal{R}(\cdot)$ are replaced with time-dependent parameters $\mu^t$ and networks $ \text{prox}_\mathcal{R}(\cdot, \alpha^t, \beta^t, t) $, respectively. 
\begin{table*}[!t]
    \scriptsize
    \centering
    \setlength{\tabcolsep}{2.5pt}
    \renewcommand{\arraystretch}{0.7}
    \caption{$\spadesuit$: Shared $\mathcal{R}(\cdot)$ weights, $\clubsuit$: Unshared $\mathcal{R}(\cdot)$ weights. Quantitative results on the coronal PD datasets using equispaced undersampling patterns at $R=4, 6$, and $8$ with \textbf{5 and 15 unrolls}. The \colorbox{best}{{\bf best}} and \colorbox{2ndbest}{second-best} values for each architecture are highlighted along with their relative difference to \textbf{10 unrolls}.}
    
    \begin{tabular}{@{}p{0.2cm}rcccccc|ccccc@{}}
        \toprule
         &  &  & \multicolumn{5}{c}{{\bf U-Net}} & \multicolumn{5}{c}{{\bf ResNet}}\\
         \cmidrule(lr){4-8} \cmidrule(lr){9-13}
        \textbf{ } & \textbf{R} & & \textbf{\tiny VS ($\spadesuit$)} & \shortstack{\textbf{\tiny VS ($\clubsuit$)} \\ \tiny (Fine-tuned)} & \textbf{\tiny ADMM ($\spadesuit$)} & \shortstack{\textbf{\tiny ADMM ($\clubsuit$)} \\ \tiny (Fine-tuned)} & \textbf{\tiny Ours} & \textbf{\tiny VS ($\spadesuit$)} & \shortstack{\textbf{\tiny VS ($\clubsuit$)} \\ \tiny (Fine-tuned)} & \textbf{\tiny ADMM ($\spadesuit$)} & \shortstack{\textbf{\tiny ADMM ($\clubsuit$)} \\ \tiny (Fine-tuned)} & \textbf{\tiny Ours}\\ \midrule
        
        \multirow{6}{*}{\rotatebox{90}{\parbox{1.7cm}{\centering \textbf{5 unrolls}}}}
        & \multirow{2}{*}{$\times 4$} & PSNR$\uparrow$ 
        & 40.58 & 40.71 & 40.86 & \secondbest{40.93} & \best{40.94}
        & 40.66 & 41.19 & 41.16 & \secondbest{41.38} & \best{41.41} \\  
        
        &  & SSIM$\uparrow$ 
        & 0.963 & 0.963 & 0.964 & \best{0.964} & \secondbest{0.964}
        & 0.963 & 0.963 & 0.964 & \secondbest{0.965} & \best{0.966} \\  
        \arrayrulecolor{gray} \cmidrule(lr){2-13}
        
        & \multirow{2}{*}{$\times 6$} & PSNR$\uparrow$ 
        & 38.01 & 38.57 & 38.49 & \secondbest{38.92} & \best{39.08}
        & 39.25 & 39.57 & 39.44 & \best{39.76} & \secondbest{39.65}    \\   
        
        &  & SSIM$\uparrow$ 
        & 0.943 & 0.948 & 0.947 & \secondbest{0.951} & \best{0.952}
        & 0.952 & 0.952 & 0.954 & \best{0.955} & \secondbest{0.954}   \\   
        \arrayrulecolor{gray} \cmidrule(lr){2-13}
        
        & \multirow{2}{*}{$\times 8$} & PSNR$\uparrow$ 
        & 35.63 & 35.81 & 35.79 & \secondbest{35.99} & \best{36.45}
        & 35.94 & 36.44 & 36.40 & \best{36.65} & \secondbest{36.76}  \\   
        
        &  & SSIM$\uparrow$ 
        & 0.916 & 0.918 & 0.917 & \secondbest{0.920} & \best{0.925}
        & 0.919 & 0.922 & 0.924 & \secondbest{0.925} & \best{0.925}   \\  
        \arrayrulecolor{gray} \cmidrule(lr){2-13}

        \multirow{6}{*}{\rotatebox{90}{\parbox{1.7cm}{\centering \textbf{15 unrolls}}}}
        & \multirow{2}{*}{$\times 4$} & PSNR$\uparrow$ 
        & 40.36 & 40.84 & 40.59 & \best{41.01} & \secondbest{40.88}
        & 41.16 & 41.36 & 41.44 & \secondbest{41.48} & \best{41.52}  \\  
        
        &  & SSIM$\uparrow$ 
        & 0.962 & 0.964 & 0.963 & \best{0.965} & \secondbest{0.964}
        & 0.964 & 0.964 & \secondbest{0.966} & 0.965 & \best{0.966}   \\   
        \arrayrulecolor{gray} \cmidrule(lr){2-13}

        & \multirow{2}{*}{$\times 6$} & PSNR$\uparrow$ 
        & 38.08 & 38.77 & 38.74 & \best{39.01} & \secondbest{38.94}
        & 39.61 & 39.78 & 39.61 & \secondbest{39.80} & \best{39.83}    \\   
        
        &  & SSIM$\uparrow$ 
        & 0.944 & 0.950 & 0.948 & \best{0.952} & \secondbest{0.951}
        & \best{0.955} & 0.953 & 0.952 & 0.954 & \secondbest{0.954}   \\   
        \arrayrulecolor{gray} \cmidrule(lr){2-13}
        
        & \multirow{2}{*}{$\times 8$} & PSNR$\uparrow$ 
        & 35.44 & 35.97 & 36.25 & \secondbest{36.50} & \best{36.51}
        & 36.41 & 36.87 & 36.95 & \secondbest{37.09} & \best{37.09}  \\   
        
        &  & SSIM$\uparrow$ 
        & 0.908 & 0.918 & 0.923 & \best{0.926} & \secondbest{0.926}
        & 0.923 & 0.926 & 0.928 & \secondbest{0.927} & \best{0.929}   \\   
                
        \arrayrulecolor{black} \bottomrule
    \end{tabular}
\label{tab:5_15unrolls_quantitative}
\end{table*}

\figref{fig:appdx_stage_wise_results} presents examples of how the reconstruction evolves when comparing shared VSQP and ADMM with its time-embedded counterparts. The shared methods exhibit persistent errors over iterations that the proximal operator fails to eliminate. However, integrating our proposed time-embedding methods into the baseline algorithms effectively addresses these issues, demonstrating gradual denoising, as intended. 

\begin{figure}[t]
\centering
\begin{minipage}{0.50\textwidth}
  \centering
  \begin{table}[H]
    \scriptsize
    \centering
    \setlength{\tabcolsep}{3.5pt}
    \renewcommand{\arraystretch}{0.7}
    \caption{The comparison results for different model sizes ($T=10$, $R=4$, and Coronal PD).}
    \vspace{+0.2cm}
    \begin{tabular}{@{}lr|c|c|c|c@{}}
        \toprule
        & \textbf{Method} & \textbf{Channel} & \textbf{$\#$ Param.} & \textbf{PSNR$\uparrow$ } & \textbf{SSIM$\uparrow$ } \\ \midrule

        \multirow{6}{*}{\rotatebox{90}{\parbox{1.5cm}{\centering \textbf{ResNet-VSQP}}}}
        & \textbf{Shared $\mathcal{R}$} & 64 & 592,129 & 41.11 & 0.965  \\
        & \textbf{Unshared $\mathcal{R}$} & 64 & 5,921,281 & 40.99 & 0.963  \\
        \arrayrulecolor{gray} \cmidrule(lr){2-6}

        & \textbf{Shared $\mathcal{R}$} & 96 & 1,330,561 & 41.09 & 0.965  \\
        & \textbf{Unshared $\mathcal{R}$} & 96 & 13,305,601 & 41.00 & 0.963  \\
        \arrayrulecolor{gray} \cmidrule(lr){2-6}
        
        & \textbf{Ours ($T=5$)} & 64 & 866,571 & \textbf{41.41} & \textbf{0.966}  \\
        & \textbf{Ours ($T=10$)} & 64 & 866,581 & \textbf{41.43} & \textbf{0.965}  \\

        \toprule

        & \textbf{Method} & \textbf{Channel} & \textbf{$\#$ Param.} & \textbf{PSNR$\uparrow$ } & \textbf{SSIM$\uparrow$ } \\ \midrule
        
        \multirow{6}{*}{\rotatebox{90}{\parbox{1.5cm}{\centering \textbf{UNet-VSQP}}}}
        & \textbf{Shared $\mathcal{R}$} & [32, 64, 128] & 1,724,035 & 40.50 & 0.962  \\
        & \textbf{Unshared $\mathcal{R}$} & [32, 64, 128] & 17,240,341 & 40.31 & 0.960  \\
        \arrayrulecolor{gray} \cmidrule(lr){2-6}

        & \textbf{Shared $\mathcal{R}$} & [64, 128, 256] & 6,878,467 & 40.77 & 0.964  \\
        & \textbf{Unshared $\mathcal{R}$} & [64, 128, 256] & 68,784,661 & 40.55 & 0.962  \\
        \arrayrulecolor{gray} \cmidrule(lr){2-6}
        
        & \textbf{Ours ($T=5$)} & [32, 64, 128] & 1,963,469 & \textbf{40.94} & \textbf{0.964}  \\
        & \textbf{Ours ($T=10$)} & [32, 64, 128] & 1,963,479 & \textbf{40.99} & \textbf{0.964}  \\

        \arrayrulecolor{black} \bottomrule
        
    \end{tabular}
\label{tab:ablation_number_param}
\vspace{-0.1cm}
\end{table}
\end{minipage}\hfill
\begin{minipage}{0.47\textwidth}
  \begin{minipage}{\textwidth}
    \centering
    \begin{table}[H]
    \scriptsize
    \centering
    \setlength{\tabcolsep}{3.5pt}
    \renewcommand{\arraystretch}{1.0}
    \caption{The comparison results for time-embedded unrolling networks with different time-embedding hyperparameters ($T=10$, $R=6$, and Coronal PDFS).}
    \vspace{+0.1cm}
    \begin{tabular}{@{}c|c|c|c|c@{}}
        \toprule
        \textbf{Freq.} & \textbf{Emb. dim.} & \textbf{Hidden layer dim.} & \textbf{PSNR$\uparrow$ } & \textbf{SSIM$\uparrow$ } \\ \midrule

        1,000 & \multirow{3}{*}{32} & \multirow{3}{*}{128} & 34.34 & 0.824  \\
        5,000 & & & 34.35 & 0.822  \\
        10,000 & & & $\textbf{34.44}$ & $\textbf{0.825}$  \\
        \arrayrulecolor{gray} \cmidrule(lr){1-5}

        \multirow{3}{*}{10,000} & 32 & \multirow{3}{*}{128} & $\textbf{34.44}$ & $\textbf{0.825}$  \\
        & 64 & & 34.32 & 0.824  \\
        & 96 & & 34.34 & 0.824  \\
        \arrayrulecolor{gray} \cmidrule(lr){1-5}
        
        \multirow{3}{*}{10,000} & \multirow{3}{*}{32} & 64 & 34.38 & 0.824  \\
        &  & 128 & $\textbf{34.44}$ & $\textbf{0.825}$  \\
        &  & 196 & 34.37 & 0.824  \\

        \arrayrulecolor{black} \bottomrule
    \end{tabular}
\label{tab:time_emb_hyperparameter}
\vspace{-0.1cm}
\end{table}

    \vspace{+0.05em} 
  \end{minipage}  
\end{minipage}
\end{figure}

\section{Additional Details on the Ablation Studies} \label{sec:supp_detail_ablation}
This section presents further implementation details and results for the experiments described in~\secref{sec:4_5}.

\paragraph{Robust Time-Embedded Unrolling with Different Numbers of Unrolls} 
Time-embedding denoisers can recognize temporal sequence information, allowing them to adaptively apply varying degrees of denoising at different stages, as shown in \figref{fig:appdx_stage_wise_results}. Based on these observations, we hypothesized that our proposed method can achieve stable performance even with a reduced or increased number of unrolling iterations. We compared our approach with both shared and unshared unrolling methods, where each was trained with $T =$ 5 and 15 unrolls. In this experiment, the unshared networks were fine-tuned to improve performance; for a detailed rationale, please refer to~\appref{sec:unshared_overfitting}.~\tabref{tab:5_15unrolls_quantitative} presents quantitative reconstruction results for $T\!=\!$ 5 and 15 unrolls. With fewer iterations ($T\!=\!$ 5 unrolls), our approach exhibits greater flexibility and robustness compared to the shared baseline algorithms, which experience performance degradation as the number of unrolls decreases. Notably, for $R=8$, both shared and unshared baselines for each architecture show significant PSNR degradation when using $T\!=\!$ 5 unrolls. In contrast, our proposed method maintains performance even with fewer iterations, showing either a slight improvement or only minimal degradation compared to $T\!=\!$ 10 unrolls, depending on the choice of the proximal operator architecture. Moreover, as the number of iterations are increased ($T\!=\!$ 15 unrolls), our proposed method maintains its robustness and consistently improves performance against the baseline models with shared $\mathcal{R}(\cdot)$, while introducing only a minimal increase in computational complexity.

\begin{figure*}
    \centering
    \includegraphics[width=0.9\linewidth]{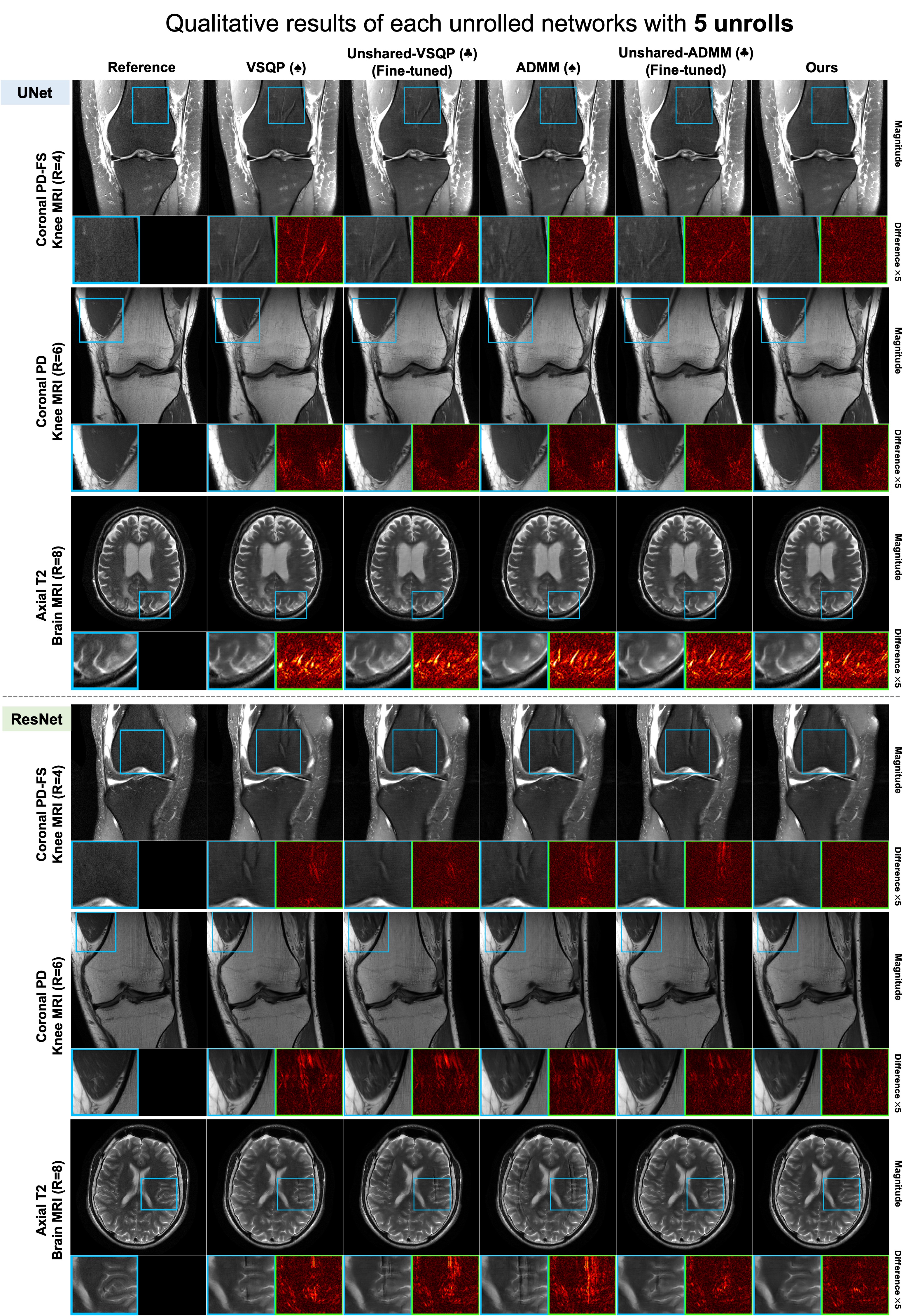} 
    \caption{$\spadesuit$: Shared $\mathcal{R}(\cdot)$ weights, $\clubsuit$: Unshared $\mathcal{R}(\cdot)$ weights. Qualitative comparisons of each unrolled network with \textbf{$T\!=\!$ 5 unrolls} for \textbf{U-Net} and \textbf{ResNet} proximal operators. In each proximal operator, \textbf{Top:} Results for $R=4$ using PD data. \textbf{Middle:} Results for $R=6$ using PD-FS data. \textbf{Bottom:} Results for $R=8$ using Axial T2-W data. The proposed methods still perform well with fewer iterations, effectively reducing artifacts.}
    \label{fig:appdx_results_5unroll} 
\end{figure*}

\begin{figure*}
    \centering
    \includegraphics[width=0.9\linewidth]{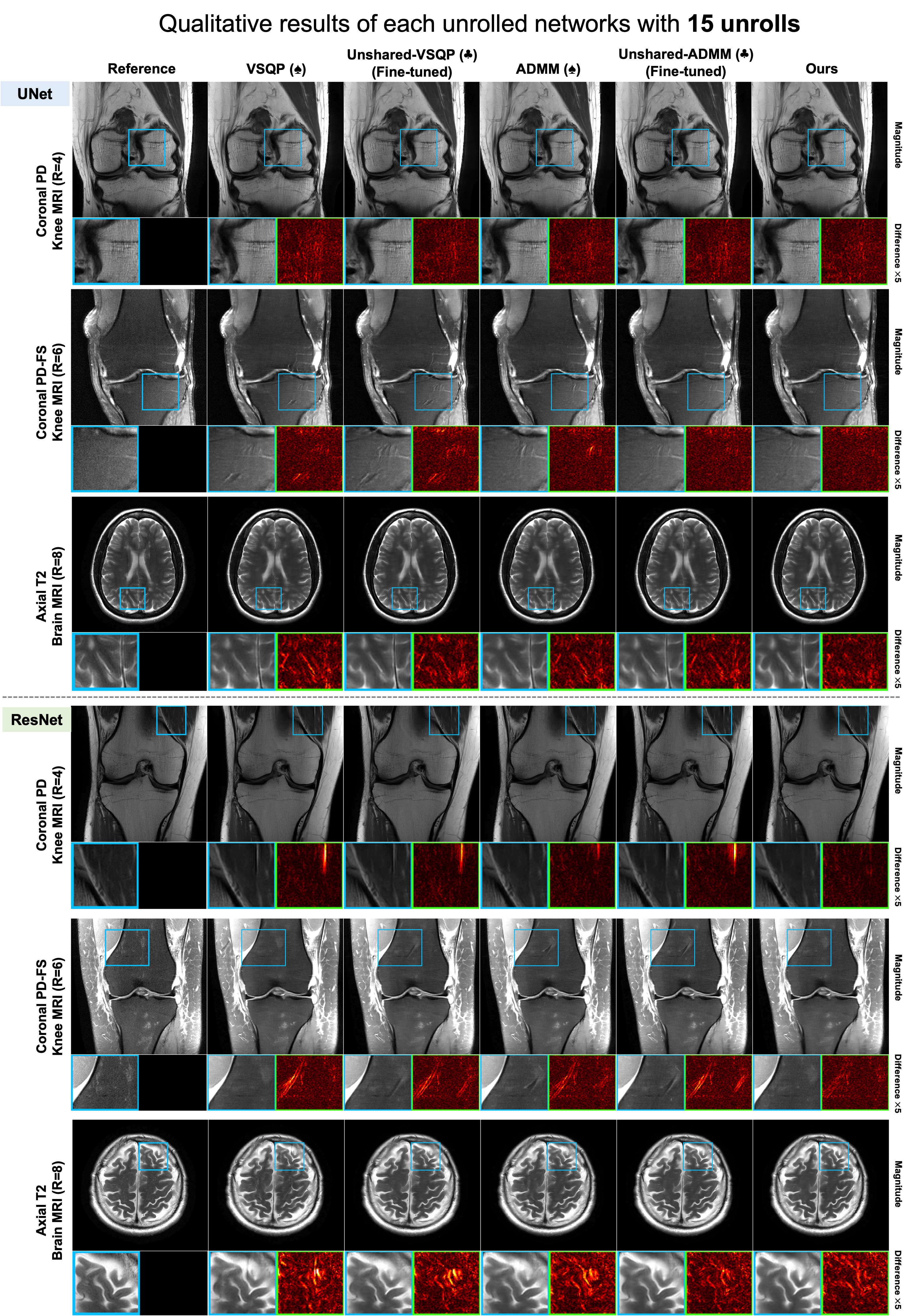} 
    \caption{$\spadesuit$: Shared $\mathcal{R}(\cdot)$ weights, $\clubsuit$: Unshared $\mathcal{R}(\cdot)$ weights. Qualitative comparisons of each unrolled network with \textbf{$T\!=\!$ 15 unrolls} for \textbf{U-Net} and \textbf{ResNet} proximal operators. In each proximal operator, \textbf{Top:} Results for $R=4$ using PD data. \textbf{Middle:} Results for $R=6$ using PD-FS data. \textbf{Bottom:} Results for $R=8$ using Axial T2-W data. The proposed methods effectively reduce artifacts and sharpen images, whereas the baseline methods fail to achieve this, even with 15 unrolls.}
    \label{fig:appdx_results_15unroll} 
\end{figure*}

The qualitative results in \figref{fig:appdx_results_5unroll} and \figref{fig:appdx_results_15unroll} for $T\!=\!$ 5 and 15 unroll iterations, respectively, support these quantitative observations. Our proposed method effectively reduces artifacts and enhances image sharpness, while the shared and unshared baseline models struggle to achieve similar improvements, both with fewer and increased iterations.

\begin{table}[t]
    \scriptsize
    \centering
    \setlength{\tabcolsep}{5.0pt}
    \renewcommand{\arraystretch}{0.8}    
    \caption{$\spadesuit$: Shared $\mathcal{R}(\cdot)$ weights, $\clubsuit$: Unshared $\mathcal{R}(\cdot)$ weights. Quantitative results are reported on the Coronal PD, Coronal PD-FS, and axial T2 datasets, with non-uniform undersampling patterns at acceleration rates $R=4$, $6$, and $8$. The \textbf{best} result for each architecture are highlighted.}
    \vspace{+0.2cm}
    \begin{tabular}{@{}ccc|cc|cc|c@{}}
        \toprule
         & \textbf{R} & & \textbf{VSQP ($\spadesuit$)} & \textbf{VSQP ($\clubsuit$)} & \textbf{ADMM ($\spadesuit$)} & \textbf{ADMM ($\clubsuit$)} & \textbf{Ours} \\ \midrule
        
        \multirow{6}{*}{\rotatebox{90}{\parbox{1.5cm}{\centering \textbf{Cor. PD}}}}
        & \multirow{2}{*}{$\times 4$} 
        & PSNR$\uparrow$ & 40.10 & 40.26 & 40.20 & 40.13 & \textbf{40.43} \\
        & & SSIM$\uparrow$ & 0.961 & 0.961 & 0.961 & 0.961 & \textbf{0.962} \\
        \arrayrulecolor{gray} \cmidrule(lr){2-8}

        & \multirow{2}{*}{$\times 6$} 
        & PSNR$\uparrow$ & 38.41 & 38.59 & 38.64 & 38.72 & \textbf{38.73} \\
        & & SSIM$\uparrow$ & 0.947 & 0.947 & 0.948 & 0.949 & \textbf{0.949} \\
        \arrayrulecolor{gray} \cmidrule(lr){2-8}

        &\multirow{2}{*}{$\times 8$} 
        & PSNR$\uparrow$ & 37.30 & 37.39 & 37.64 & 37.52 & \textbf{37.76} \\
        & & SSIM$\uparrow$ & 0.935 & 0.934 & \textbf{0.939} & 0.938 & 0.938 \\

        \toprule
         & \textbf{R} & & \textbf{VSQP ($\spadesuit$)} & \textbf{VSQP ($\clubsuit$)} & \textbf{ADMM ($\spadesuit$)} & \textbf{ADMM ($\clubsuit$)} & \textbf{Ours} \\ \midrule
        
        \multirow{6}{*}{\rotatebox{90}{\parbox{1.5cm}{\centering \textbf{Cor. PDFS}}}}
        & \multirow{2}{*}{$\times 4$} 
        & PSNR$\uparrow$ & 35.61 & 35.60 & 35.62 & 35.66 & \textbf{35.68} \\
        & & SSIM$\uparrow$ & 0.855 & 0.853 & 0.853 & \textbf{0.856} & 0.854 \\
        \arrayrulecolor{gray} \cmidrule(lr){2-8}

        & \multirow{2}{*}{$\times 6$} 
        & PSNR$\uparrow$ & 34.59 & 34.55 & 34.70 & 34.61 & \textbf{34.74} \\
        & & SSIM$\uparrow$ & 0.828 & 0.824 & 0.827 & 0.827 & \textbf{0.830} \\
        \arrayrulecolor{gray} \cmidrule(lr){2-8}

        & \multirow{2}{*}{$\times 8$} 
        & PSNR$\uparrow$ & 33.76 & 33.97 & 34.06 & 34.07 & \textbf{34.14} \\
        & & SSIM$\uparrow$ & 0.809 & 0.808 & 0.810 & 0.810 & \textbf{0.812} \\

        \toprule
        & \textbf{R} & & \textbf{VSQP ($\spadesuit$)} & \textbf{VSQP ($\clubsuit$)} & \textbf{ADMM ($\spadesuit$)} & \textbf{ADMM ($\clubsuit$)} & \textbf{Ours} \\ \midrule
        
        \multirow{6}{*}{\rotatebox{90}{\parbox{1.5cm}{\centering \textbf{Axial T2}}}}
        & \multirow{2}{*}{$\times 4$} 
        & PSNR$\uparrow$ & 35.90 & 36.36 & 36.50 & 36.42 & \textbf{36.52} \\
        & & SSIM$\uparrow$ & \textbf{0.932} & 0.931 & 0.930 & 0.931 & 0.930 \\
        \arrayrulecolor{gray} \cmidrule(lr){2-8}

        & \multirow{2}{*}{$\times 6$} 
        & PSNR$\uparrow$ & 35.03 & 35.11 & \textbf{35.26} & 35.25 & 35.21 \\
        & & SSIM$\uparrow$ & \textbf{0.917} & 0.916 & 0.915 & 0.905 & 0.914 \\
        \arrayrulecolor{gray} \cmidrule(lr){2-8}

        & \multirow{2}{*}{$\times 8$} 
        & PSNR$\uparrow$ & 34.09 & 34.31 & 34.37 & 34.39 & \textbf{34.51} \\
        & & SSIM$\uparrow$ & 0.908 & 0.908 & 0.903 & \textbf{0.906} & 0.904 \\

        \arrayrulecolor{black} \bottomrule
    \end{tabular}
\label{tab:non_uniform_mask}
\vspace{-0.3cm}
\end{table}

\paragraph{Efficiency Relative to the Number of Parameters and Time-Embedding Hyperparameters} To assess efficiency with respect to the number of parameters, we explored the effect of increasing the number of parameters on performance, which resulted in higher total parameter counts in both the shared and unshared baselines. As shown in~\tabref{tab:ablation_number_param}, we use the following setups: (1) increasing the channels in ResNet residual blocks from 64 to 96, (2) increasing the channels in U-Net up/downsampling blocks from [32, 64, 128] to [64, 128, 256], and (3) using $T=10$, $R=4$, with the Coronal PD dataset.

For efficiency relative to the time-embedding hyperparameters, we evaluated (1) the frequency of the sinusoidal encoding, (2) the embedding dimension, and (3) the number of hidden channels in the MLP layers. The experiments were conducted using a U-Net architecture with $T = 10$ and $R = 6$ on the Coronal PDFS dataset. Implementation details and results are provided in~\tabref{tab:time_emb_hyperparameter}.

\section{Details on the Extended Experiments} \label{sec:supp_extended_exp}
This section provides additional implementation details and results for the experiments described in~\secref{sec:extended_exp}.

\paragraph{Experiments on Non-Uniform Undersampling Masks} 

As shown in~\tabref{tab:non_uniform_mask}, our method consistently outperforms the baselines in terms of PSNR in all cases except for the Axial T2 dataset at R=6. For SSIM, our method shows improvement in most cases for the PD and PD-FS datasets, although no improvement is observed for the Axial T2 dataset.

\paragraph{Comparison with Diffusion Model-Based Reconstruction} 
\begin{table}[h]
    \scriptsize
    \centering
    \setlength{\tabcolsep}{5.5pt}
    \caption{The comparison results with diffusion-based model (DDS).}
    \vspace{+0.2cm}
    \renewcommand{\arraystretch}{1.1}
    \begin{tabular}{@{}r|c|c|c|c@{}}
        \toprule
        \textbf{Method} & \textbf{Data} & \textbf{R} & \textbf{PSNR$\uparrow$ } & \textbf{SSIM$\uparrow$ } \\ \midrule

        \textbf{DDS (100)} & \textbf{PD} & $\times$4 & 37.41{\tiny $\pm$3.25} & 0.940{\tiny $\pm$0.029}  \\
        \textbf{Ours (U-Net)} & \textbf{PD} & $\times$4 & \textbf{40.09}{\tiny $\pm$2.51} & \textbf{0.958}{\tiny $\pm$0.017}  \\

        \arrayrulecolor{black} \bottomrule
        
    \end{tabular}
    \label{tab:dds_comparison}
\vspace{-0.1cm}
\end{table}

Since DDS requires 320$\times$320 inputs due to its generative pre-trained prior, we additionally evaluated PSNR and SSIM using the central 320$\times$320 region. Note that this differs from the results reported in~\tabref{tab:table_quant_main}, where evaluations were performed on images of size 320$\times$368, aligned with the original raw k-space data. We set the number of sampling steps to 100 for DDS. For our method, we used $T=10$ unrolls. All experiments were conducted on the Coronal PD test dataset with an acceleration factor of $R=4$. As shown in~\tabref{tab:dds_comparison}, our method outperforms DDS in both PSNR and SSIM. Furthermore, diffusion-based reconstruction requires tens to hundreds of neural function evaluations (NFEs) during inference~\cite{chung2024decomposed}, which remains far from the efficiency needed for large-scale or real-time applications. In contrast, our time-embedded unrolled networks achieve more promising results with substantially fewer NFEs (e.g., 5–10), even with a smaller network architecture compared to diffusion-based models.

\section{Additional Qualitative Results} \label{sec:additional_results}

\begin{figure*}[h]
    \centering
    \includegraphics[width=0.97\linewidth]{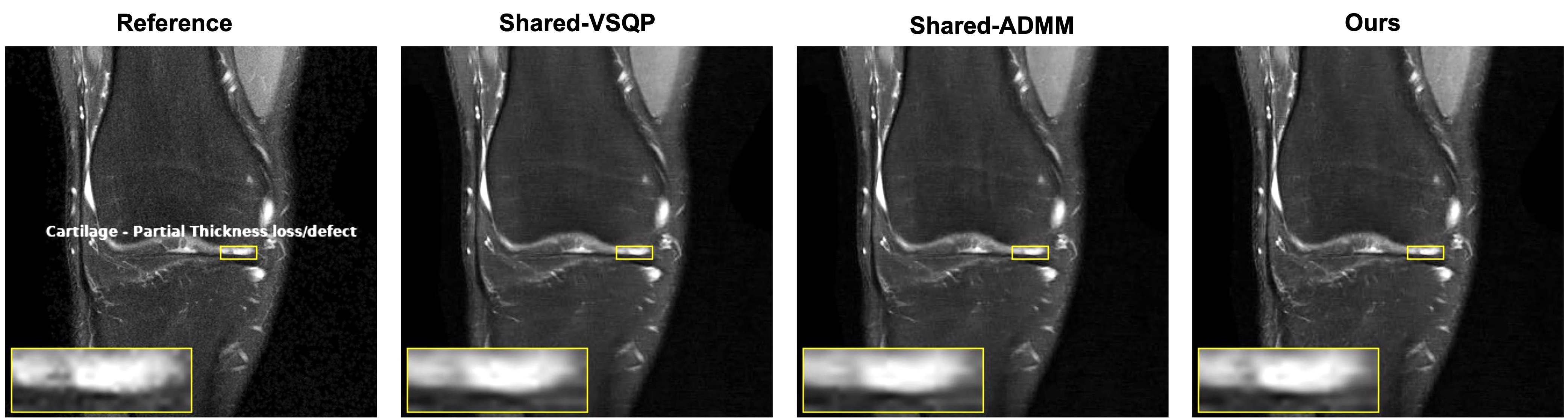} 
    \caption{Reconstruction results with annotated pathological regions.}
    \label{fig:appdx_fastmri_plus} 
\end{figure*}

\paragraph{Pathological Region Inspection Using fastMRI+}
We leveraged the annotations of pathological regions provided by fastMRI+~\cite{zhao2022fastmri_plus} to further validate the strengths of our method. As shown in~\figref{fig:appdx_fastmri_plus}, our approach produces clearer contrast in the pathological regions compared to other methods, which is further corroborated with radiologist assessments, as detailed next. 

\paragraph{Radiologist Readings }
For a subset of all the data processed in this study, a musculoskeletal radiologist with over 30 years of experience blindly reviewed the reconstructed images from the different methods. The radiologist’s assessments highlight improvements achieved by our method that are critical for diagnostic purposes. Details are provided below.

\begin{itemize}[leftmargin=*]
\item \figref{fig:result_unet} (Middle) exhibits aliasing artifacts in the distal femoral metaphysis medially on PD-weighted images for VSQP, Unshared-VSQP, ADMM, Unshared-ADMMs. The artifacts are effectively removed in the proposed method (ours).

\item \figref{fig:result_unet} (Bottom) shows blurring in the right occipital lobe in VSQP, Unshared-VSQP, ADMM, Unshared-ADMMM. This is notably improved in the proposed method, where the gyri and sulci appear sharper.

\item \figref{fig:result_resnet} (Top) shows visible aliasing artifacts in the central aspect of the distal femoral condyle of the inset for VSQP, Unshared-VSQP, ADMM, and Unshared-ADMM. This artifact is removed in the proposed method. The prominent penetrating intraosseous vessel at the lateral aspect of the proximal tibia (lower left on the image) appears sharper in proposed method, though overall image sharpness is similar among methods.

\item \figref{fig:result_resnet} (Middle): There is an oblong, hypointense appearing  aliasing artifact on the non-fat saturated PD-weighted images of the knee joint, just distal to the posteromedial femoral condyle, seen on VSQP, Unshared-VSQP, ADMM, and Unshared-ADMM, when compared to the reference. This artifact is removed from the image for the proposed method. Thus, only the proposed method accurately resembles the reference image. 

\item \figref{fig:result_resnet} (Bottom) reveals aliasing artifacts in VSQP, Unshared-VSQP, ADMM, Unshared-ADMM, depicted as curvilinear, oblique hypointense signal in the occipital lobe. The artifact is nearly completely removed in Unshared-ADMM, and only vaguely seen. The artifact is completely removed in ours, most accurately resembling the reference image.

\item In \figref{fig:appdx_fastmri_plus}, on the reference image there is a focal area of T2-hyperintense signal, most consistent with a partial/full thickness cartilage defect in the anteromedial trochlea. Images reconstructed by VSQP, Unshared-VSQP, ADMM, and Unshared-ADMMM reveal significant blurring in this area. Proposed method shows the least amount of blurring compared to other methods, and shows the hyperintense region in the trochlear articular cartilage with the most fidelity compared to the reference data.
\end{itemize}

\begin{figure*}
    \centering
    \includegraphics[width=0.9\linewidth]{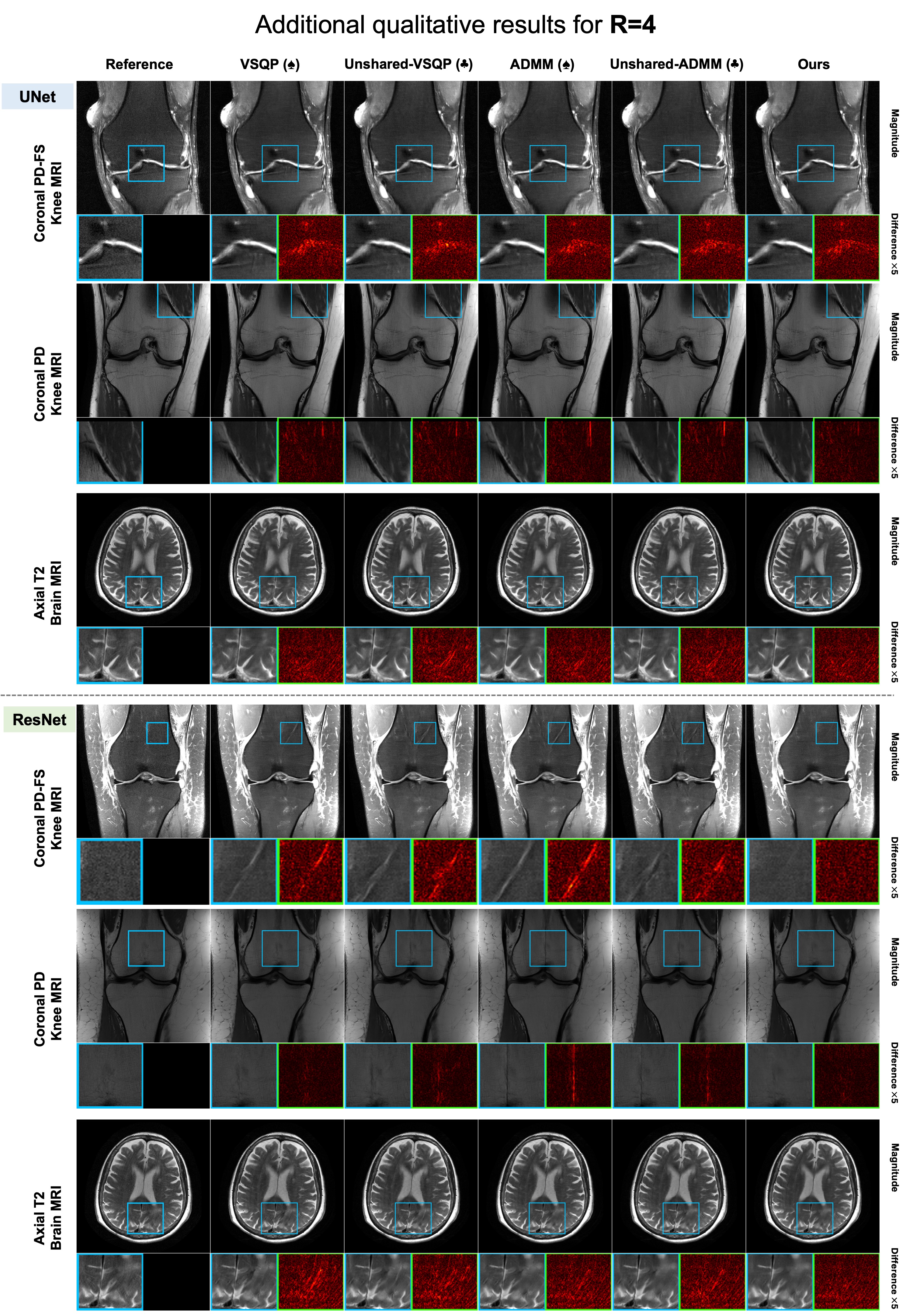} 
\caption{$\spadesuit$: Shared $\mathcal{R}(\cdot)$ weights, $\clubsuit$: Unshared $\mathcal{R}(\cdot)$ weights. Qualitative comparisons for \(\mathbf{R = 4}\) across datasets for each proximal operator (\textbf{$T\!=\!$ 10 unrolls}). Our proposed method consistently demonstrates superior performance by reducing artifacts.}
    \label{fig:appdx_results_R4}
\end{figure*}

\begin{figure*}
    \centering
    \includegraphics[width=0.9\linewidth]{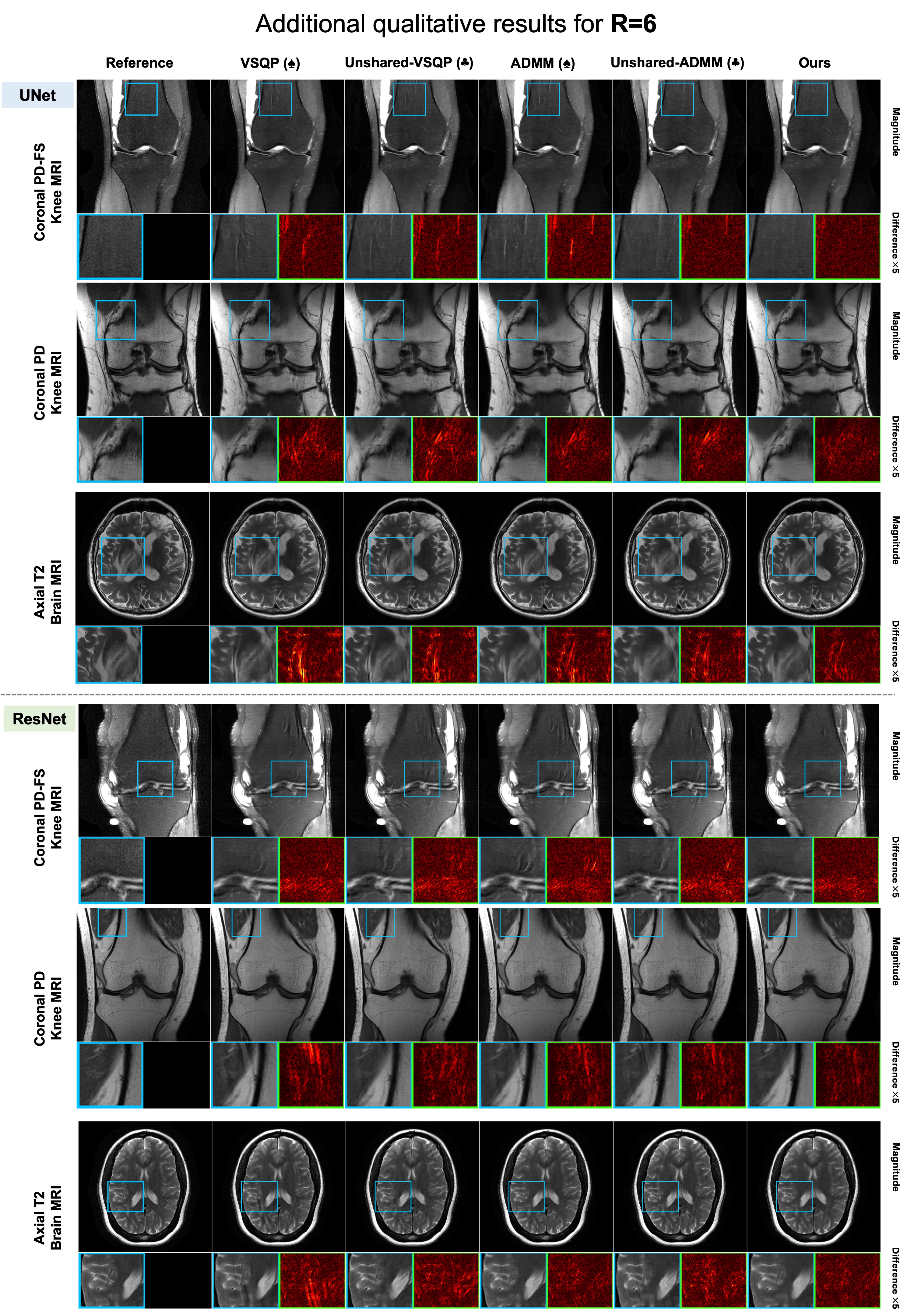} 
    \caption{$\spadesuit$: Shared $\mathcal{R}(\cdot )$ weights, $\clubsuit$: Unshared $\mathcal{R}(\cdot)$ weights. Qualitative comparisons for \(\mathbf{R = 6}\) across datasets for each proximal operator (\textbf{$T\!=\!$ 10 unrolls}). Our proposed method consistently demonstrates superior performance by reducing artifacts. Furthermore, it enhances image sharpness, as shown in the results for the axial T2 data with both U-Net and ResNet proximal operators.}
    \label{fig:appdx_results_R6}
\end{figure*}

\begin{figure*}
    \centering
    \includegraphics[width=0.9\linewidth]{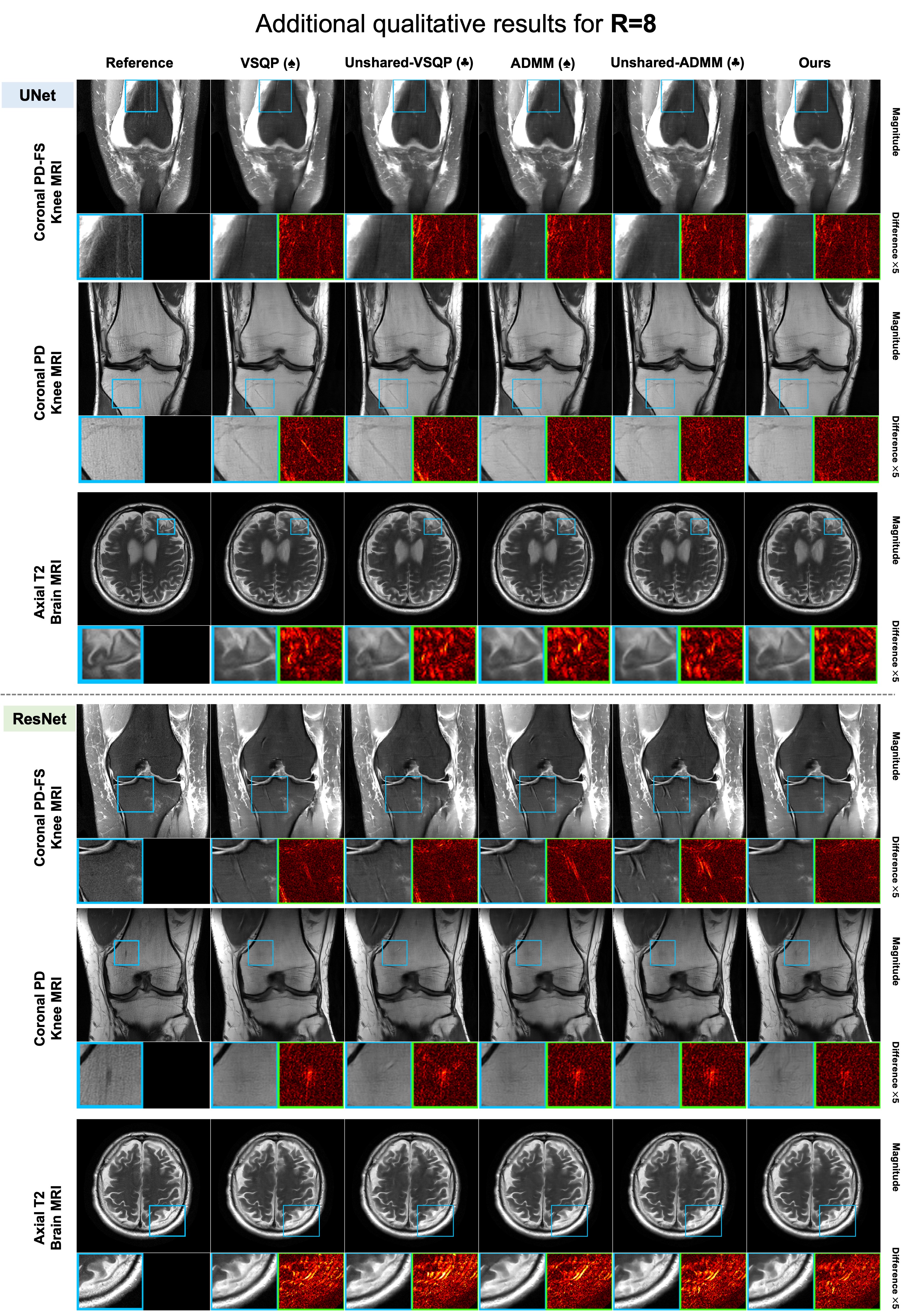} 
    \caption{$\spadesuit$: Shared $\mathcal{R}(\cdot)$ weights, $\clubsuit$: Unshared $\mathcal{R}(\cdot)$ weights. Qualitative comparisons for \(\mathbf{R = 8}\) across datasets for each proximal operator (\textbf{$T\!=\!$ 10 unrolls}). Similar to R=4 and R=6, R=8 also demonstrates artifact reduction and image sharpening. Through \figref{fig:appdx_results_R4} to \figref{fig:appdx_results_R8}, our proposed method shows superior performance across all configurations.}
    \label{fig:appdx_results_R8}
\end{figure*}

\paragraph{Additional Qualitative Examples}
We provide additional representative reconstruction examples that demonstrate the visual superiority of our proposed method, since PSNR/SSIM do not necessarily align with perception, as discussed in \secref{sec:4_3}. \figref{fig:appdx_results_R4}, \figref{fig:appdx_results_R6}, and \figref{fig:appdx_results_R8} shows reconstruction results across all datasets and proximal operator architectures for \(R = 4\), \(6\), and \(8\), respectively, using the implementations described in the main text.

\section{Extended Quantitative Results with Standard Deviation} \label{sec:std}
\tabref{tab:table_quant_main_std} summarizes the standard deviation of PSNR and SSIM for the same settings as in \tabref{tab:table_quant_main}.

\begin{table*}[!t]
    \scriptsize
    \centering
    \setlength{\tabcolsep}{2.0pt}
    \renewcommand{\arraystretch}{0.8} 
    \caption{$\spadesuit$: Shared $\mathcal{R}(\cdot)$ weights; $\clubsuit$: Unshared $\mathcal{R}(\cdot)$ weights. Standard deviations of PSNR and SSIM results for the same settings in \tabref{tab:table_quant_main}.}
    \begin{tabular}{@{}p{0.2cm}rcccccc|ccccc@{}}
        \toprule
         &  &  & \multicolumn{5}{c}{{\bf U-Net}} & \multicolumn{5}{c}{{\bf ResNet}}\\
         \cmidrule(lr){4-8} \cmidrule(lr){9-13}
        \textbf{ } & \textbf{R} & & \textbf{\tiny VSQP ($\spadesuit$)} & \textbf{\tiny VSQP ($\clubsuit$)} & \textbf{\tiny ADMM ($\spadesuit$)} & \textbf{\tiny ADMM ($\clubsuit$)} & \textbf{\tiny Ours} & \textbf{\tiny VSQP ($\spadesuit$)} & \textbf{\tiny VSQP ($\clubsuit$)} & \textbf{\tiny ADMM ($\spadesuit$)} & \textbf{\tiny ADMM ($\clubsuit$)} & \textbf{\tiny Ours}\\ \midrule
        \multirow{6}{*}{\rotatebox{90}{\parbox{1.7cm}{\centering \textbf{Coronal PD}}}}
        & \multirow{2}{*}{$\times 4$} & PSNR 
        &  2.56 &  2.59 &  2.42 &  2.41 &  2.40
        &  2.95 &  3.04 &  2.97 &  3.03 &  2.73  \\  
        
        &  & SSIM 
        &  0.02 &  0.02 &  0.02 &  0.01 &  0.01
        &  0.02 &  0.02 &  0.02 &  0.02 &  0.02   \\   
        \arrayrulecolor{gray} \cmidrule(lr){2-13}

        & \multirow{2}{*}{$\times 6$} & PSNR 
        &  2.10 &  2.31 &  2.21 &  2.10 &  2.17
        &  2.77 &  2.81 &  2.74 &  2.74 &  2.38  \\   
        
        &  & SSIM 
        &  0.02 &  0.03 &  0.02 &  0.02 &  0.02 
        &  0.02 &  0.02 &  0.02 &  0.02 &  0.02   \\   
        \arrayrulecolor{gray} \cmidrule(lr){2-13}

        & \multirow{2}{*}{$\times 8$} & PSNR 
        &  2.15 &  2.29 &  2.22 &  2.09 &  2.00 
        &  2.60 &  2.70 &  2.66 &  2.67 &  2.28  \\

        &  & SSIM 
        &  0.03 &  0.04 &  0.04 &  0.03 &  0.03 
        &  0.04 &  0.04 &  0.04 &  0.04 &  0.03   \\  
        \arrayrulecolor{black} \midrule

        \multirow{6}{*}{\rotatebox{90}{\parbox{1.8cm}{\centering \textbf{Coronal PD-FS}}}}
        & \multirow{2}{*}{$\times 4$} & PSNR   
        &  2.73 &  2.79 &  2.80 &  2.77 &  2.76 
        &  2.78 &  2.80 &  2.82 &  2.79 &  2.88  \\  
                
        &  & SSIM 
        &  0.10 &  0.10 &  0.10 &  0.10 &  0.09 
        &  0.10 &  0.10 &  0.10 &  0.10 &  0.10  \\  
        \arrayrulecolor{gray} \cmidrule(lr){2-13}

        & \multirow{2}{*}{$\times 6$} & PSNR 
        &  2.56 &  2.61 &  2.59 &  2.53 &  2.58 
        &  2.69 &  2.70 &  2.70 &  2.68 &  2.75   \\

        &  & SSIM 
        &  0.11 &  0.11 &  0.11 &  0.11 &  0.11 
        &  0.11 &  0.11 &  0.11 &  0.11 &  0.11   \\  
        \arrayrulecolor{gray} \cmidrule(lr){2-13}
        
        & \multirow{2}{*}{$\times 8$} & PSNR 
        &  2.41 &  2.44 &  2.43 &  2.36 &  2.44 
        &  2.46 &  2.48 &  2.53 &  2.54 &  2.60   \\

        &  & SSIM 
        &  0.11 &  0.11 &  0.11 &  0.11 &  0.11 
        &  0.12 &  0.12 &  0.12 &  0.12 &  0.12   \\  
        
        \arrayrulecolor{black} \midrule

        \multirow{6}{*}{\rotatebox{90}{\parbox{1.8cm}{\centering \textbf{Axial T2-W}}}}
        & \multirow{2}{*}{$\times 4$} & PSNR   
        &  3.01 &  2.88 &  2.99 &  2.91 &  2.92 
        &  3.18 &  3.17 &  3.29 &  3.12 &  3.19 \\

        &  & SSIM 
        &  0.06 &  0.05 &  0.05 &  0.05 &  0.06
        &  0.06 &  0.06 &  0.06 &  0.06 &  0.06   \\  
        \arrayrulecolor{gray} \cmidrule(lr){2-13}

        & \multirow{2}{*}{$\times 6$} & PSNR 
        &  3.05 &  2.68 &  2.88 &  2.79 &  2.98
        &  3.04 &  2.99 &  3.22 &  3.08 &  3.10 \\

        &  & SSIM 
        &  0.08 &  0.06 &  0.06 &  0.06 &  0.06
        &  0.07 &  0.07 &  0.07 &  0.07 &  0.06   \\  
        \arrayrulecolor{gray} \cmidrule(lr){2-13}

        & \multirow{2}{*}{$\times 8$} & PSNR 
        &  2.40 &  2.35 &  2.55 &  2.57 &  2.61 
        &  2.60 &  2.53 &  2.77 &  2.55 &  2.82 \\

        &  & SSIM 
        &  0.06 &  0.06 &  0.06 &  0.07 &  0.07
        &  0.07 &  0.07 &  0.07 &  0.07 &  0.07  \\

        \arrayrulecolor{black} \bottomrule
    \end{tabular}

\label{tab:table_quant_main_std}
\vspace{-0.3cm}
\end{table*}


\section{Discussions} \label{sec:discuss}

\paragraph{Second-Moment Matching.}

Since our methods are inspired by the VAMP framework, an assessment of second-moment matching for VAMP is desirable. Second-moment matching is typically evaluated by examining the estimated variances $ \upsilon_x^{t}$ and $ \upsilon_z^{t} $ across iterations, as in~\eqref{eq:lmsse_Onsager} and~\eqref{eq:vamp2}. However, even if consistent trends are empirically observed in these estimates, this would not constitute a formal proof. This is further complicated in our case, as we hypothesize that a time-embedded neural network models all update steps in~\eqref{eq:vamp1}-\eqref{eq:vamp2}, and the learnable scalar parameter $\rho^t$ encapsulates the entire process described in~\eqref{eq:lmsse_Onsager}. As a result, $\upsilon_x^{t}$ and $\upsilon_z^{t}$ are embedded within black-box neural modules and are not explicitly accessible.

Instead, to indirectly assess whether second-moment matching holds, we analyzed the relationships between intermediate estimates. Specifically, we examined the empirical differences between $\mathbf{x}^t$ and $\mathbf{u}^t$, and between $\mathbf{u}^t $ and $\mathbf{r}^{t}$ in~\algoref{algo:alg_proposed}, as these pairs are intrinsically related to $\upsilon_x^{t}$ and $\upsilon_z^{t}$, respectively. If these differences remained consistently small across iterations, it provided evidence that the underlying variance estimates are stable. The empirical difference between $\mathbf{x}^t$ and the reconstructed $\mathbf{u}^t$ was reported in~\appref{sec:supp_x_r_comparison}. As shown in~\tabref{tab:table_x_u_comparison}, the difference (normalized MSE) ranges from $1.01 \times 10^{-9}$ to $3.35 \times 10^{-2}$, demonstrating stable behavior. These findings suggest that second-moment matching is empirically preserved, despite the use of learned components.

\paragraph{Lipschitz Constant or Gradient Explosion/Vanishing of Time-embedding (FiLM) Layers.} Consider a network where at each time step $t \in \{1, \ldots, T\}$, there are $K$ consecutive layers within the proximal operator networks composed of intermediate transformations followed by FiLM modulation. 
For each layer $k \in \{1, \ldots, K\}$,
\begin{equation}
x^{(t,k)} = \mathrm{FiLM}( f_{k}(x^{(t,k-1)}), t ),
\end{equation}
where $f_{k}(\cdot)$ denotes the intermediate layers (e.g., convolution + activation) preceding the FiLM block at layer $k$.
Suppose each intermediate layer is Lipschitz continuous with constant $L_{k}$, and satisfies
\begin{equation}
\|f_{k}(x)\| \leq L_{k} \|x\| + \delta_{k},
\end{equation}
for some small $\delta_{k} \geq 0$, allowing for nonzero bias or offset when $f_{k}(0) \neq 0$. Similarly,  each FiLM block is Lipschitz continuous with constant $ \Gamma_{t,k} $, satisfying
\begin{equation}
\|\mathrm{FiLM}(z, t)\| \leq \Gamma_{t,k} \|z\| + B_{t,k}.
\end{equation}
The composite function then satisfies:
\begin{equation}
\|\mathrm{FiLM}(f_{k}(x), t)\| \leq \Gamma_{t,k} L_{k} \|x\| + (\Gamma_{t,k} \delta_{k} + B_{t,k}).
\end{equation}
Thus, the overall Lipschitz constant of the composite layer at layer $k$ and time $t$ is $\Gamma_{t,k} L_{k}$. 
As $T$ grows, if each $\Gamma_{t,k} L_{k}$ is strictly less than 1, their product decays exponentially, which may cause vanishing activations and hinder learning. If any are $\geq$ 1, the product can grow exponentially, causing exploding activations and instability. Thus, controlling cumulative constant, $\prod_{t=1}^T \prod_{k=1}^K \Gamma_{t,k} L_{k}$ is crucial for stable training. 

While a formal proof is not provided, as we do not analytically characterize the Lipschitz constants of individual components, our empirical results support the stability of the proposed approach. In particular, we adopt the U-Net architecture and FiLM modules commonly used in diffusion denoising tasks. In the diffusion model literature, a large number of diffusion steps ($T \geq 1,000$) is typically employed, which has been shown to promote stable training. In our unrolled networks, where a substantially smaller number of steps is used (\eg $T=5$--$15$), we observe that training remains stable, suggesting that the reduced $T$ does not compromise empirical stability. These observations underscore the empirical nature of our Lipschitz constant bounds and their relevance to practical performance.

\end{document}